\documentclass[aps,prd,preprint,showpacs,nofootinbib]{revtex4}
\usepackage{epsfig}
\usepackage{amsmath}
\usepackage{amssymb}
\usepackage{amsbsy}
\usepackage{amsfonts}
\usepackage{graphics}
\usepackage{latexsym}
\usepackage{mathrsfs}
\usepackage{dcolumn}
\usepackage{wasysym}
\usepackage[dvips]{color}
\usepackage{graphicx}
\usepackage[figuresright]{rotating}

\begin{document}
\vspace*{-1.5cm}

\begin{raggedright}
DESY 09-157\\[-0.75em]
ITEP-LAT-2009-13\\[-0.75em]
Edinburgh 2009/12\\[-0.75em]
LTH 844
\end{raggedright}
\vspace*{-0.25cm}

\title{Probing the finite temperature phase transition with $N_f=2$
  nonperturbatively improved Wilson fermions}

\author{V.G.~Bornyakov,$^{1,2}$ R.~Horsley,$^{3}$ S.M.~Morozov,$^{2}$
Y.~Nakamura,$^{4,5}$ M.I.~Polikarpov,$^{2}$ P.E.L.~Rakow,$^{6}$
G.~Schierholz,$^{7}$ and T.~Suzuki$^{8,5}$}

\affiliation{\vspace*{0.3cm}
$^1$ Institute for High Energy Physics IHEP, 142281 Protvino, Russia\\
$^2$ Institute of Theoretical and
Experimental Physics ITEP, 117259 Moscow, Russia \\
$^3$ School of Physics and Astronomy, University of Edinburgh, Edinburgh
EH9 3JZ, UK\\
$^4$ Institut f\"ur Theoretische Physik, Universit\"at Regensburg,
93040 Regensburg, Germany\\
$^5$ RIKEN, Radiation Laboratory, Wako 351-0158, Japan \\
$^6$ Theoretical Physics Division, Department of Mathematical Sciences,
University of Liverpool, Liverpool L69 3BX, UK\\
$^7$ Deutsches Elektronen-Synchrotron DESY, 22603 Hamburg, Germany \\
$^8$ Institute for Theoretical Physics,
Kanazawa University, Kanazawa 920-1192, Japan \\
\vspace*{-0.75cm}
}
\author{QCDSF--DIK Collaboration}




\begin{abstract}

The critical temperature and the nature of the QCD finite temperature phase
transition are determined for $N_f=2$ dynamical flavors of nonperturbatively
improved Wilson fermions. The calculations are performed on large
lattices with temporal extents $N_t=12$, $10$ and $8$, and lattice spacings down
to $a = 0.075 \, \mbox{fm}$. We find the deconfinement and chiral phase
transitions to take place at the same temperature. Our results are in broad
agreement with a second order phase transition in the chiral
limit. The critical temperature at the physical quark mass is found to
be $T_c = 174(3)(6)\, \mbox{MeV}$.

\end{abstract}


\maketitle

\section{Introduction}

One of the basic questions in finite temperature QCD is: What is
the nature of the finite temperature phase transition, and at which
temperature does it happen? In spite of enormous computational
efforts~\cite{BB,W}, the answer to this question has remained 
controversial. The Wuppertal group~\cite{W} finds the deconfining
transition, which we identify with the peak of the Polyakov-loop
susceptibility, and the chiral transition at widely separated
temperatures. In contrast, the Brookhaven/Bielefeld collaboration claims both
temperatures to  
coincide. Furthermore, the Brookhaven/Bielefeld collaboration~\cite{BB2}
quotes a transition temperature of $T_c = 196(3) \, \mbox{MeV}$, 
while the transition temperatures found by the Wuppertal
group~\cite{W2} are $T_c = 170(7)\, \mbox{MeV}$ for the deconfining
transition and $T_c = 146(5)\, \mbox{MeV}$ for the chiral
transition. Both groups use rooted staggered fermions, but with
different levels of improvement. The Brookhaven/Bielefeld collaboration uses
{\it asqtad} as well as {\it p4} fermions, while the Wuppertal group
employs twice 
iterated stout smeared links in the fermion matrix. It has been
argued~\cite{W} that the discrepancy is largely due to the rather
coarse lattices used by the Brookhaven/Bielefeld collaboration. Indeed, 
inititial calculations of this group were limited to lattices of temporal
extent $N_t \leq 6$, corresponding to lattice spacings of $a \gtrsim
0.2\, \mbox{fm}$, while the Wuppertal group performed simulations on
lattices of extent $N_t=10$, $8$ and $6$, and attempted a continuum
extrapolation. More recently, the Brookhaven/Bielefeld collaboration
has extended their calculations to lattices of temporal extent $N_t =
8$~\cite{BBB} and found that with decreasing lattice spacing $T_c$
shifted by $5 - 7 \, \mbox{MeV}$ towards smaller values.

The connection between deconfining and chiral transition has been the
subject of several phenomenological considerations. Naively, one would
expect the temperature of the deconfinement transition to lie below that of
the chiral transition, if different at all. This turns out to
be the case, for example, in the Polyakov-loop extended Nambu--Jona-Lasinio
model~\cite{Weise}. More likely is that both transitions
occur at the same temperature, as Polyakov loop and chiral condensate mix
at finite dynamical quark masses. The consequence would be a simultaneous
enhancement of both the chiral and Polyakov-loop susceptibilities along the
transition line~\cite{mix,mix2,mix3,mix4}. 

To clarify the issue, an independent investigation of the nature of
the finite temperature phase transition is needed. In this work we
shall perform simulations with $N_f=2$ dynamical flavors of
nonperturbatively $O(a)$ improved Wilson fermions, so-called clover
fermions, and plaquette gauge action on lattices of temporal extent
$N_t=12$, $10$ and $8$.  This action has been successfully employed by
the ALPHA, UKQCD, QCDSF and CERN/Rome collaborations in calculations
at zero temperature, on whose results we can draw.
Besides that, clover fermions have an exact flavor symmetry, which we
consider a big advantage over staggered fermions, as the nature of the
finite temperature phase transition largely depends on the flavor
degrees of freedom. A disadvantage though is the lack of chiral symmetry.
Preliminary results of our work have been reported in~\cite{DIK}.  

The paper is organized as follows. In sec.~II we present the action,
and in sec.~III we introduce the order parameters, which are the main focus
of the analysis. The lattice data are presented in sec.~IV and in the
Appendix. In sec.~V we recapitulate the equation of state, that is
expected to describe the thermodynamic properties of QCD with two dynamical
flavors near the critical point. Relations between second derivatives
of the partition function are called Maxwell relations. One such
relation connects the chiral susceptibility to the derivative of the
plaquette with respect to mass, and will be derived in sec.~VI. In
sec.~VII we compute the transition temperature from the Polyakov-loop
susceptibility, the chiral susceptibility and the correlator of Polyakov
loop and chiral condensate, and fit its dependence on the quark mass
to the prediction of the equation of state. Finally, in sec.~VIII we conclude.

\section{Lattice simulation}

We consider mass-degenerate sea quarks. The fermionic action for each
of the two flavors reads
\begin{equation}
\begin{split}
 S_F =  a^4 \sum_x \,\Big\{  
 &\frac{1}{2 a} \,\sum_\mu \,\bar{\psi}(x) \;
 U_\mu(x)\, \left[\gamma_\mu - 1\right]\, \psi(x +a \hat{\mu}) \\
 -\; &\frac{1}{2 a}\, \sum_\mu \,\bar{\psi}(x) \;
 U^\dagger_\mu(x -a \hat{\mu})\, \left[\gamma_\mu + 1\right]\, \psi(x
 -a \hat{\mu})\\  
 -\; &c_{SW} \,\frac{i}{2a} \, \sum_{\mu\nu} \, 
 \bar{\psi}(x)\,\sigma_{\mu\nu}\, P_{\mu\nu}(x)\, \psi(x) 
  + (m + m_c) \, \bar{\psi}(x)\, \psi(x) \Big\} 
 \label{SW_action} 
\end{split}
\end{equation}  
 where $P_{\mu\nu}$ is the clover-leaf form of the
 lattice field strength tensor, 
\begin{equation}
\begin{split}
P_{\mu\nu}(x)=\frac{1}{4}\, \Big[& U_\nu(x)\, U_\mu(x +a \hat{\nu})\,
U_\nu^\dagger(x +a \hat{\mu})\, U_\mu^\dagger(x)\\
+&U_\nu^\dagger(x -a \hat{\nu})\, U_\mu^\dagger(x -a \hat{\mu} -a \hat{\nu})\,
U_\nu(x -a \hat{\mu} -a \hat{\nu})\, U_\mu(x -a \hat{\mu})\\
-&U_\nu(x)\, U_\mu^\dagger(x -a \hat{\mu} +a \hat{\nu})\, U_\nu^\dagger(x
-a \hat{\mu})\, U_\mu(x -a \hat{\mu})  \\
-&U_\nu^\dagger(x -a \hat{\nu})\, U_\mu(x -a \hat{\nu})\,
U_\nu(x +a \hat{\mu} -a \hat{\nu})\, U_\mu^\dagger(x)\Big] \,,
\end{split}
\end{equation}
and
 \begin{equation} 
 a m_c = \frac{1}{2 \kappa_c} \,, \quad a m = \frac{1}{2 \kappa} -
 \frac{1}{2 \kappa_c} \,  
 \end{equation} 
$\kappa_c$ being the critical value of the hopping parameter. The improvement
coefficient $c_{SW}$ was determined nonperturbatively~\cite{Rainer}. 

\begin{table}[t]
\begin{center}
\vspace*{0.25cm}
\begin{tabular}{|c|c|c|c|c|}\hline
$\beta$ & $c_{SW}$ & $V=N_s^3\, N_t$ & $\kappa_c$ & $r_0/a$ \\
\hline
5.20  & $2.0171$ & $16^3\, 8$\phantom{0} & $0.136050(17)$ & $5.454(58)$ \\
5.20  & $2.0171$ & $24^3\, 10$ & $0.136050(17)$ & $5.454(58)$ \\
5.25  & $1.9603$ & $16^3\, 8$\phantom{0} & $0.136273(7)$\phantom{0} & $5.880(26)$ \\
5.25  & $1.9603$ & $24^3\, 8$\phantom{0} & $0.136273(7)$\phantom{0} & $5.880(26)$ \\
5.25  & $1.9603$ & $32^3\, 12$ & $0.136273(7)$\phantom{0} & $5.880(26)$ \\
5.29  & $1.9192$ & $24^3\, 12$ & $0.136440(4)$\phantom{0}& $6.201(25)$ \\\hline 
\end{tabular}
\end{center}
\caption{Parameters of the simulation.}
\label{param}
\end{table}

We use the highly optimized HMC algorithm of the QCDSF
collaboration~\cite{QCDSF} for updating the gauge field. Considerable 
speedups were obtained by applying mass preconditoning \`{a} la
Hasenbusch~\cite{Hasenbusch} and putting the pseudofermion action on
multiple time scales~\cite{Sexton}. The algorithm runs about three times
faster than an equally optimized RHMC algorithm for $N_f=2+1$ dynamical
flavors~\cite{Roger}, which was one of the reasons for concentrating
on $N_f=2$ flavors first. The couplings, lattice volumes and lattice
spacings covered by our simulations are listed in Table~\ref{param}. 
The scale parameters $r_0/a$ have been taken from the zero temperature
runs of the QCDSF collaboration at the corresponding couplings. They
refer to the chiral limit $\kappa = \kappa_c$. We also list the
critical hopping parameters $\kappa_c$, which we adopted from QCDSF as
well. (For recent relevant work see~\cite{QCDSF2}.) On each lattice we
have performed simulations at up to $13$ 
different $\kappa$ values placed around the transition line. The exact
values are given in Tables~\ref{tab520}--\ref{tab529} in the
Appendix. The temperature of our lattices is given by
\begin{equation}
T=\frac{1}{N_t\, a} \,.
\end{equation}

The gauge field
configurations were generated on the BlueGene/L at KEK, the RSCC
cluster at RIKEN, the MVS-50k at the Joint Computer Center (Moscow),
on the SKIF-Chebyshev at Moscow State University, as well as on the
Altix at HLRN.

\section{Order parameter}

\begin{figure}[t]
\vspace*{-1cm}
\begin{center}
\epsfig{file=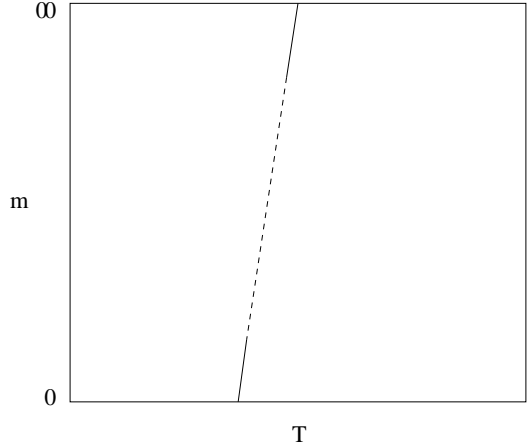,height=11.25cm,angle=270,clip=}
\end{center}
\vspace*{-1.25cm}
\caption{The phase diagram of two-flavor QCD. The solid lines indicate a
  second (bottom) and first order transition (top), respectively,
  while the dashed line 
  denotes the crossover region.}   
\label{phase}
\end{figure}

Two-flavor QCD is expected to undergo a second order transition at finite
temperature in the chiral limit and at very small quark masses. In the chiral
limit the order parameter is the subtracted scalar density
$\sigma_{-} = \sigma - \sigma_0$, defined through
\begin{equation}
\sigma=\frac{a^3}{V}\, \sum_{x} \bar{\psi}(x) \psi(x) = Z_S^{-1}
\sigma_R + \sigma_0 \,,
\label{sig}
\end{equation}
where $\sigma_R$ is the renormalized density, and $\sigma_0$ is an
additive renormalization constant, which arises from mixing of
$\sigma$ with the vacuum due to lack of chiral symmetry.  
For heavy quark
masses close to the quenched limit, the theory is known to undergo a first
order phase transition. In that limit the order parameter is
the Polyakov loop 
\begin{equation}
L=\frac{1}{N_s^3}\,\sum_{\vec{x}}\, {\rm Re}\, L(\vec{x})\, , \,
L(\vec{x})=\frac{1}{3} \, 
{\rm Tr}\, \prod^{N_t}_{x_4=1}U_4(x) \,. 
\end{equation}
In the intermediate mass region we expect the transition to be a
crossover. This scenario is sketched in Fig.~\ref{phase}. An `order
parameter', which interpolates between the two 
limits, is 
\begin{equation}
\omega = \frac{1}{\sqrt{1+(am)^2}}\,\sigma_{-} +
\frac{am}{\sqrt{1+(am)^2}}\,L \,. 
\label{sus}
\end{equation}

The temperature of the chiral transition is, for general $m$, identified with
the peak position of the chiral susceptibility   
\begin{equation}
\chi_\sigma \equiv \langle\sigma^2\rangle_c = \langle \sigma^2\rangle - \langle
\sigma\rangle^2 \,,
\end{equation}
while the peak of the Polyakov-loop susceptibility  
\begin{equation}
\chi_L \equiv  N_s^3\, \langle L^2\rangle_c \, , \, \langle
L^2\rangle_c = \left( \langle 
  L^2\rangle - \langle L\rangle^2\right) 
\end{equation}
defines the temperature of the deconfining transition. Both,
$\langle\sigma^2\rangle_c$ and $\langle L^2\rangle_c$ can be combined
into one susceptibility, following (\ref{sus}), 
\begin{equation}
\langle \omega^2\rangle_c = \frac{1}{1+(am)^2}\, \langle \sigma^2\rangle_c +
\frac{2am}{1+(am)^2}\, \langle L \sigma\rangle_c +
\frac{(am)^2}{1+(am)^2}\, \langle L^2\rangle_c \,, 
\label{phase2}
\end{equation}
where
\begin{equation}
\langle L \sigma\rangle_c = \langle L \sigma \rangle - \langle L
\rangle\,\langle \sigma \rangle \, , 
\end{equation}
which interpolates between zero and infinite quark mass. In case the crossover
temperature is unique, all three correlators, $\langle L^2\rangle_c$,\,
$\langle L \sigma\rangle_c$\, and $\langle \sigma^2\rangle_c$,\, are
expected to peak at the same transition temperature. If, on the other
hand, the deconfining 
and chiral transitions take place at far different temperatures, we
would not expect to find a distinct peak in the correlator $\langle L 
\sigma\rangle_c$. In the connected correlators $\langle
\sigma^2\rangle_c$ and $\langle L \sigma\rangle_c$ the additive
renormalization constant $\sigma_0$ drops out, which allowed us to replace
$\sigma_{-}$ by $\sigma$.   

\begin{figure}[b]
\begin{center}
\epsfig{file=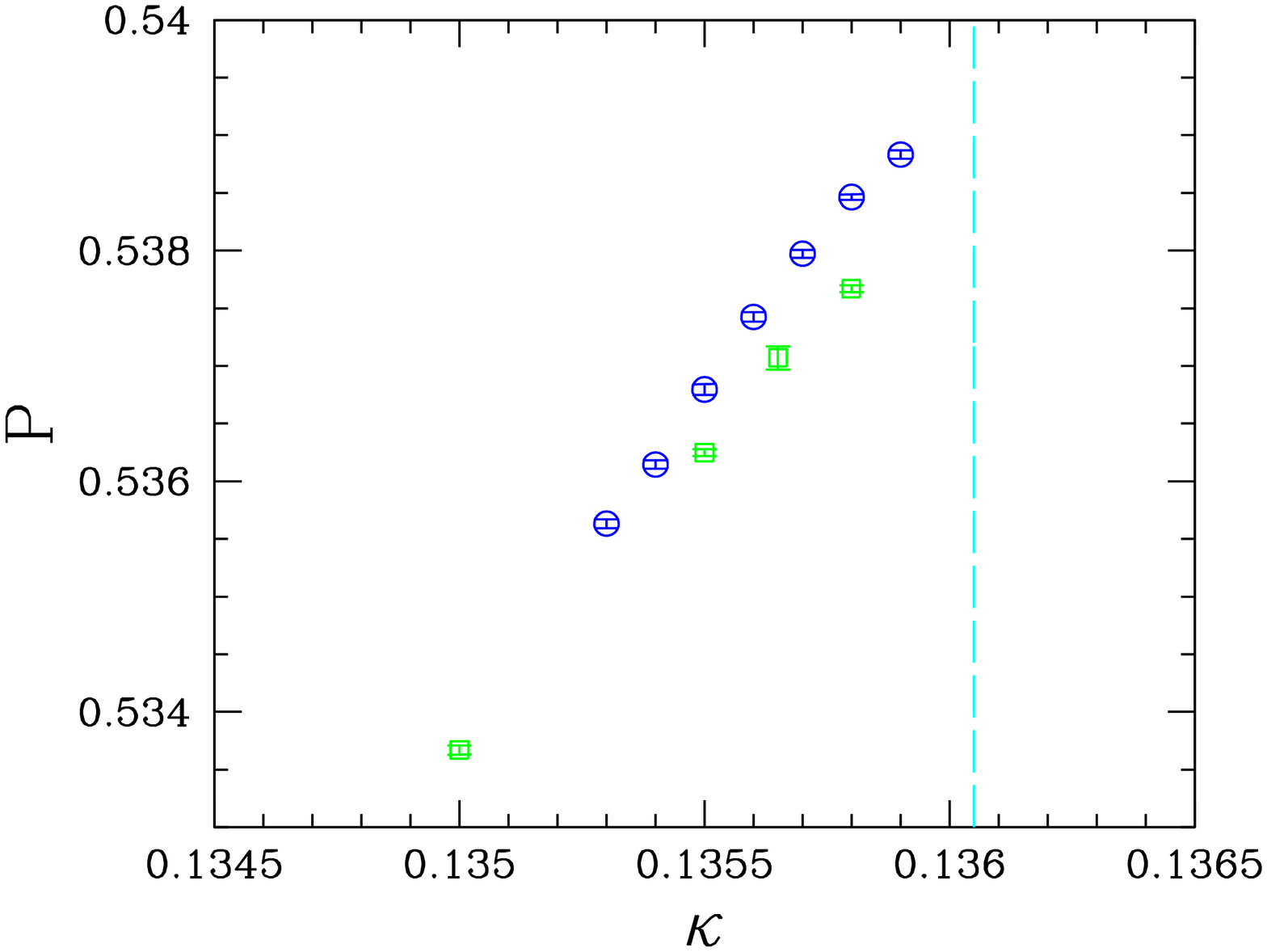,height=8cm,clip=}\\
\epsfig{file=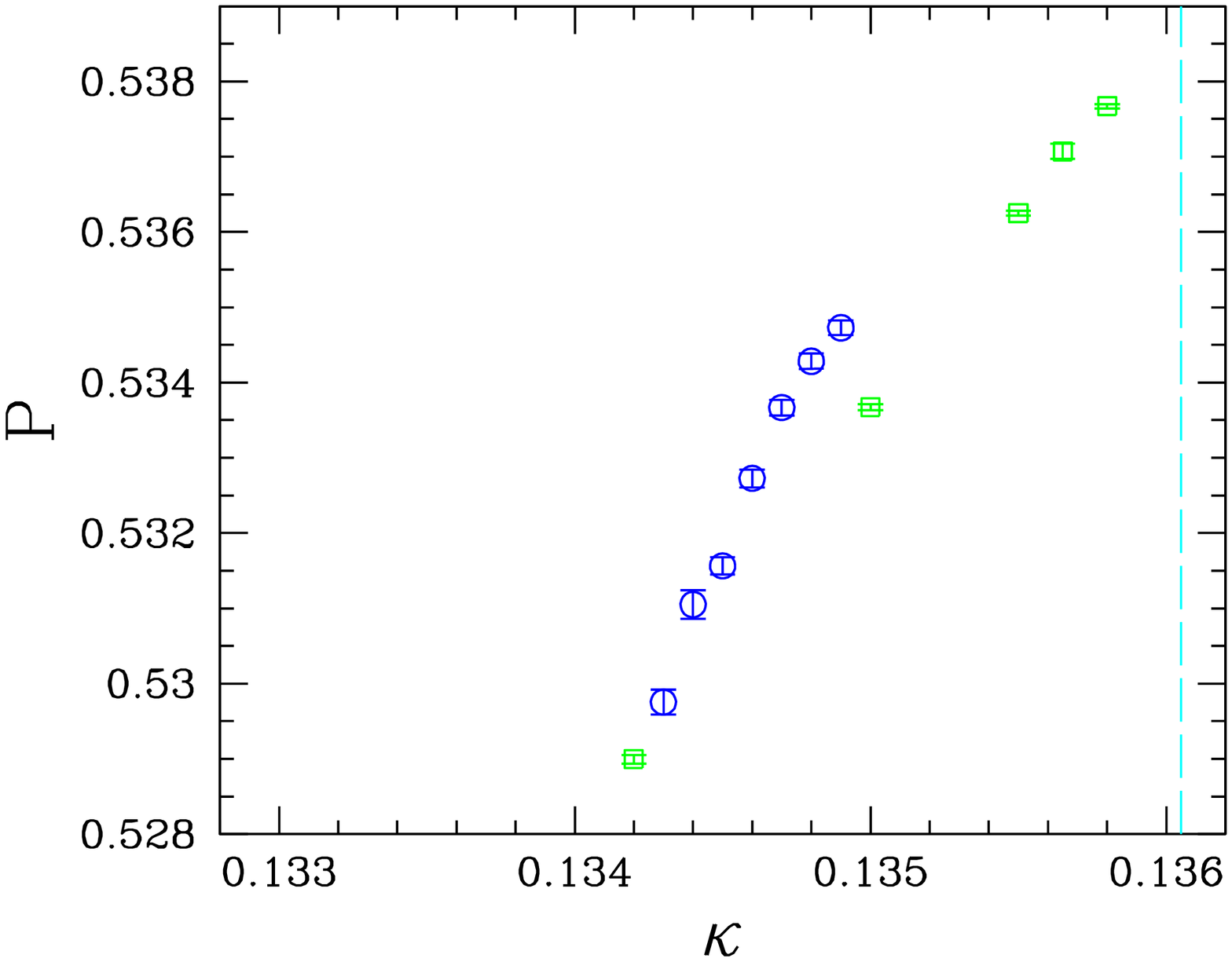,height=8cm,clip=}
\end{center}
\vspace*{-0.75cm}
\caption{The average plaquette (\textcolor{blue}{$\Circle$}) on the
  $24^3\, 10$ (top) 
  and $16^3\, 8$ lattice (bottom) at $\beta=5.20$, together with the
  average plaquette at zero temperature from the $24^3 48$ lattice
  (\textcolor{green}{$\Box$}). The dashed line indicates the position
  of $\kappa_c$.}  
\label{plaq}
\end{figure}

\begin{figure}[t]
\begin{center}
\epsfig{file=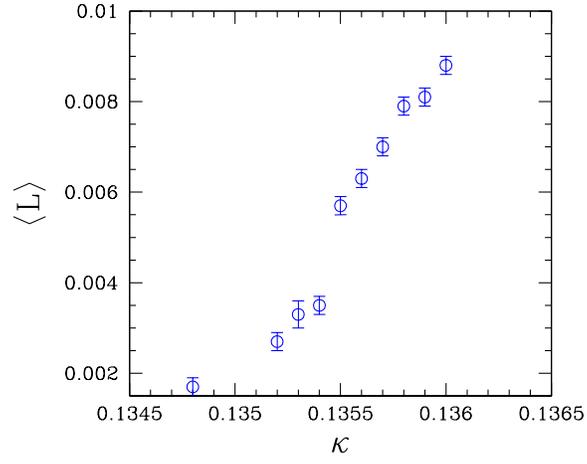,height=8cm,clip=}\\
\epsfig{file=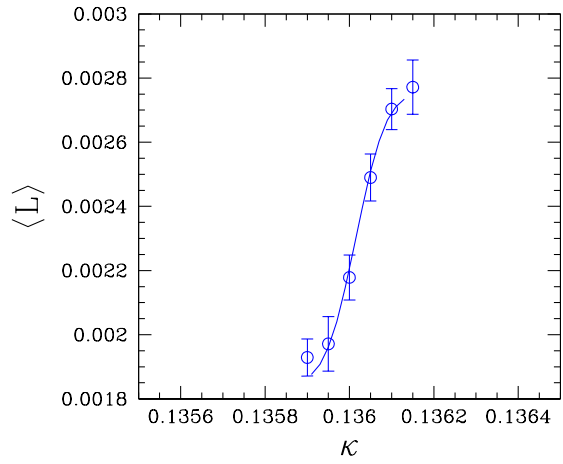,height=8cm,clip=}
\end{center}
\vspace*{-0.75cm}
\caption{The average Polyakov loop on the $24^3\, 10$ (top) 
  and $32^3\, 12$ lattice (bottom) at $\beta=5.20$ and $5.25$,
  respectively. The solid line in the bottom figure denotes the integral of
  the Gaussian in Fig.~\ref{chiL} (bottom).}    
\label{loop}
\end{figure}
\section{Data}

To compute the chiral susceptibility $\chi_\sigma$ and the
correlator of $L$ and $\sigma$, $\langle L \sigma\rangle_c$,
all we need to know is the average plaquette 
\begin{equation}
P = \frac{1}{3}\, {\rm Tr}\, \langle U_\Box \rangle \,,
\end{equation}

\clearpage
\begin{figure}[t]
\vspace*{-1.75cm}
\begin{center}
\epsfig{file=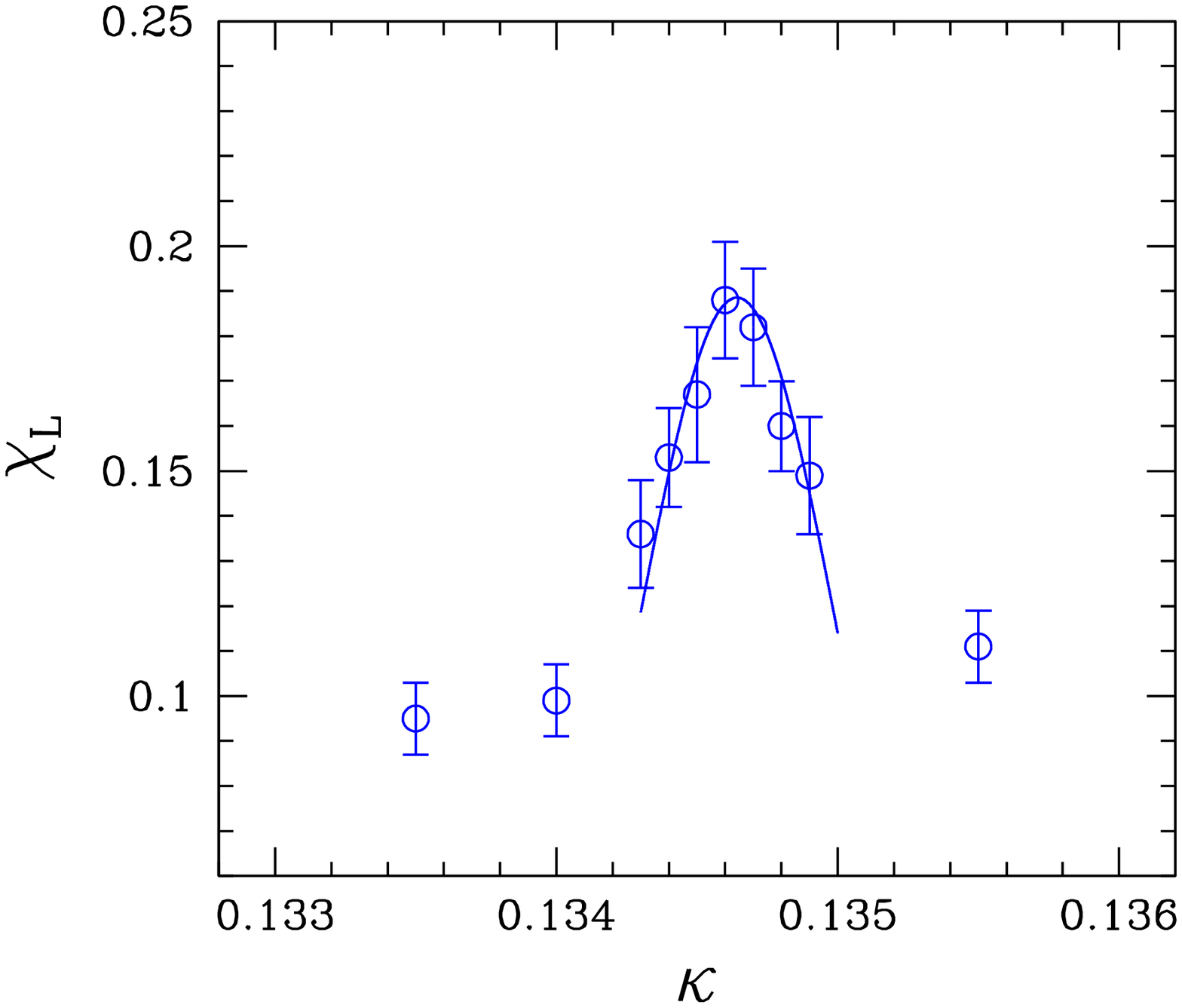,height=8cm,clip=}\\[-4em]
\epsfig{file=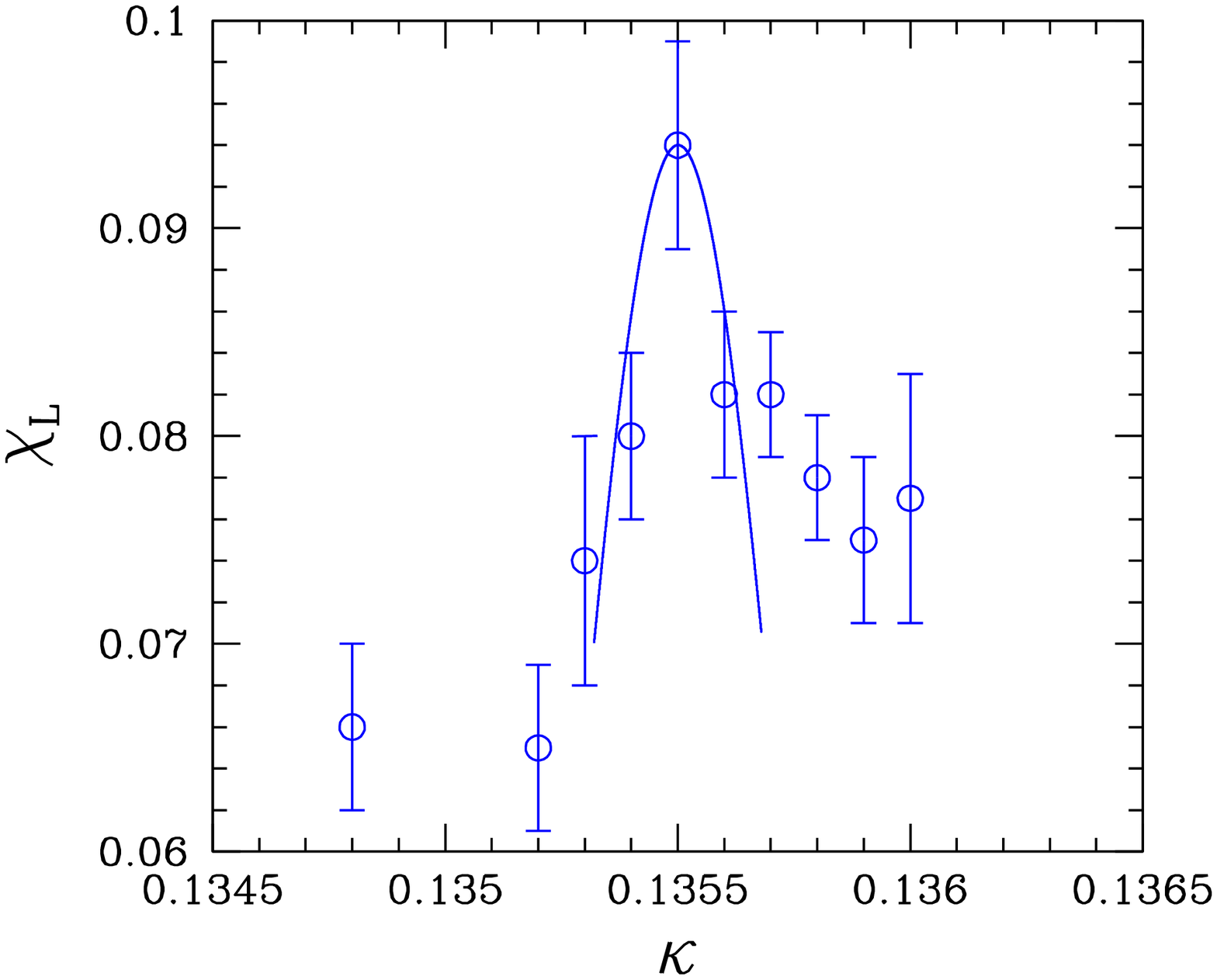,height=8cm,clip=}\\[-4em]
\epsfig{file=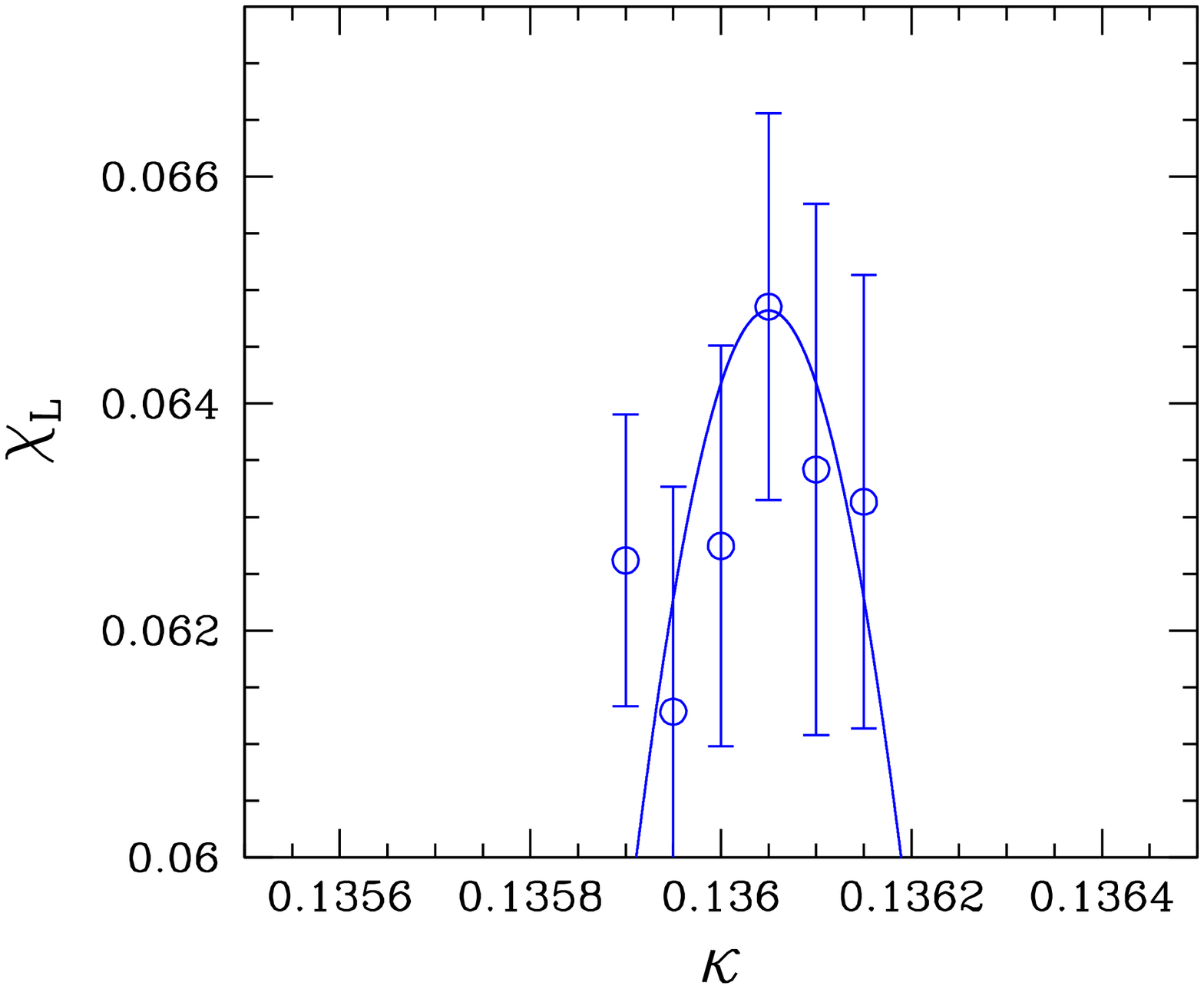,height=8cm,clip=}
\end{center}
\vspace*{-0.5cm}
\caption{The Polyakov-loop susceptibility on the $16^3\, 8$ (top), $24^3\,
  10$ (middle) and $32^3\, 12$ lattice (bottom) at $\beta=5.20$,
  $5.20$ and $5.25$, respectively, together with a Gaussian adaptation.}   
\label{susL}
\end{figure}

\clearpage
\noindent
and the average Polyakov loop $\langle L\rangle$, as we shall see.
In Tables~\ref{tab520}--\ref{tab529} in the Appendix we present our
results for $P$ and $\langle L\rangle$, as well as for the
Polyakov-loop susceptibility $\chi_L$. The statistical errors are
computed by blocked jackknife. In Fig.~\ref{plaq} we plot the
average plaquette $P$ on the $24^3\, 10$ and $16^3\, 8$ lattice at
$\beta=5.20$. For comparison, we also show
the respective numbers at zero temperature, taken from the QCDSF
collaboration. Both sets of numbers agree with each other at smaller
$\kappa$ values (corresponding to the confined phase -- see
Fig.~\ref{critical}), as expected, while at larger $\kappa$
values the finite temperature plaquette exceeds the zero temperature
one, moving closer to its perturbative value. In
Fig.~\ref{loop} we plot the average Polyakov loop $\langle L\rangle$
on the $24^3\, 10$ and $32^3\,12$ lattice at $\beta=5.20$ and $5.25$,
respectively. In both cases $\langle L\rangle$ shows a sharp increase in a 
narrow interval of $\kappa$. Finally, in Fig.~\ref{susL} we show the 
Polyakov-loop susceptibility for three different volumes and
$N_t$'s. While we observe a 
distinct peak on the $16^3\, 8$ lattice, it appears that the smaller
the quark mass is, the more is the peak washed out. This phenomena has
also been observed by the CP-PACS
group~\cite{cppacs}. (Unfortunately, the Wuppertal group does not show the
Polyakov-loop susceptibility, so we cannot compare.)  

\section{Equation of state}


We shall assume for the moment that the finite temperature transition of
two-flavor QCD is indeed of second order in the chiral limit, as indicated in
Fig.~\ref{phase}. Later on we shall see
that our data support this view, while a first order transition appears to be
unlikely. Throughout this section we shall use $\hat{\sigma}$ and
$\hat{m}$ as shorthand for the subtracted chiral condensate $\langle
\sigma_{-}\rangle$ and $am$, respectively. 

In the vicinity of the phase transition the chiral condensate, the
dynamical quark mass and the temperature are related by the equation of state 
\begin{equation} \displaystyle
\hat{m} = \hat{\sigma}^{\delta} f(t/\hat{m}^{\frac{1}{\beta\delta}}) \,,
\label{eos}
\end{equation}
where $\delta$ and $\beta$ are the critical exponents characterizing the
transition, and  
\begin{equation}
t=\frac{T-T_c(m=0)}{T_c(m=0)}\\[1em]
\end{equation}
is the reduced temperature, with $T_c(m=0)$ being the critical temperature in
the chiral limit. It is expected that the two-flavor theory is in the same 
universality class as the three-dimensional O(4) Heisenberg
model~\cite{PW}, with  
the external magnetic field and the magnetization being identified
with the bare quark mass $\hat{m}$ and the chiral condensate
$\hat{\sigma}$, respectively. The critical exponents of this model
were found to be~\cite{cex}  
\begin{equation}
\begin{split}
\frac{1}{\beta\delta} &= 0.537(7) \, , \\
\frac{1}{\delta} &= 0.206(1) \,.
\end{split}
\end{equation}

We may expect O(4) scaling to be realized in two-flavor lattice QCD 
only if the fermion action obeys flavor and chiral symmetry. Neither
of the common actions satisfy both. The problem of O(4) scaling
has been investigated by several authors (for a review see~\cite{DeTar})
with somewhat contradictory results. Simulations of two-flavor rooted
staggered fermions~\cite{Mendes} show deviations from O(4) scaling,
albeit at small $N_t$. Since staggered fermions show quite large
taste breaking effects at strong coupling, one might expect to find
O(2) critical exponents instead. This conjecture receives support
from~\cite{Kogut}. The situation improves with the adoption of {\it
  p4} fermions. In this case evidence for O(N) scaling was found just
recently~\cite{Ejiri}.    
For both Wilson~\cite{Iwasaki} and clover fermions~\cite{AAK}, on the
other hand, it has been shown that the subtracted condensate scales with O(4)
critical exponents (for $T_c(m) \geq T_c(0)$ though, see
below). In~\cite{AAK} good agreement was found even for  
pion masses up to $1\,\mbox{GeV}$. At our lattice spacings, $a \approx 0.08
\, \mbox{fm}$, we find chiral symmetry to be largely restored, which
manifests itself, for example, in the equality of the renormalization
constants of the scalar and pseudoscalar densities,
$Z_S^{Singlet}=Z_P$~\cite{GS}. So we may hope for O(4) scaling. 

\begin{figure}[b]
\vspace*{-0.75cm}
\begin{center}
\epsfig{file=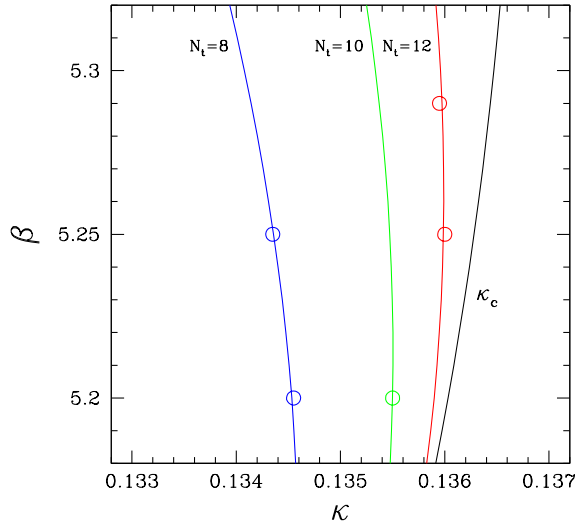,height=8cm,clip=}
\end{center}
\vspace*{-0.75cm}
\caption{The lines of critical temperature $T_c(m)$ for $N_t=8$, $10$ and $12$,
  together with the critical hopping parameter $\kappa_c$ line. The open
  circles refer to $T_c(m)$ obtained from the chiral susceptibility.} 
\label{critical}
\end{figure}

Let $T_c(m)$ denote the pseudocritical temperature at finite $m$,
which we define to 
be the temperature corresponding to the position of the peak of the chiral
susceptibility
\begin{equation}
\chi_\sigma = \frac{\partial\,\hat{\sigma}}{\partial\,\hat{m}} \,.
\end{equation}
From the scaling relation (\ref{eos}) we then derive
\begin{equation}
T_c(m)-T_c(m=0) \, \propto \, \hat{m}^{\frac{1}{\beta\delta}} \,.
\end{equation}
Assuming
\begin{equation}
m_\pi^2 \propto m \,,
\end{equation}
we thus expect to find
\begin{equation}
T_c(m)-T_c(m=0) \, \propto \, m_\pi^{1.07(1)} 
\label{o4}
\end{equation}
for a second order transition at $m=0$. A first order transition, on the other
hand, would give 
\begin{equation}
T_c(m)-T_c(m=0) \, \propto \, m_\pi^2 \,.
\label{1st}
\end{equation}

\begin{figure}[t]
\vspace*{-1.75cm}
\begin{center}
\epsfig{file=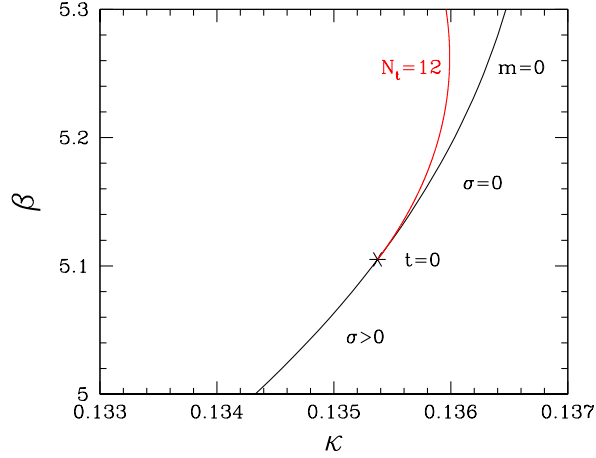,height=8cm,clip=}
\end{center}
\vspace*{-0.75cm}
\caption{The phase diagram of clover fermions for $N_t=12$.} 
\label{criticalfine}
\end{figure}

Alternatively, at fixed $N_t$, the reduced temperature $t$ can be identified 
with~\cite{Iwasaki} 
\begin{equation}
t = \beta - \beta_{\,c}^{N_t} \,,
\end{equation}
where $\beta_{\,c}^{N_t}$ is the value of $\beta$, at which the critical
temperature $T_c(m)$ line hits the critical hopping parameter $\kappa_c$ line
as we lower $\beta$. In Fig.~\ref{critical} we show the line of $T_c(m)$ 
in the $\kappa - \beta$ plane for $N_t=8$, $10$ and $12$.

Expanding the scaling function $f$ in (\ref{eos}) to lowest nontrivial
order in $t$, {\it i.e.} neglecting terms of $O(t^2)$, we obtain    
\begin{equation}
\hat{m}=A\hat{\sigma}^\delta + B\,t\,\hat{\sigma}^{\delta-\frac{1}{\beta}} 
\label{eosexp}
\end{equation} 
with $A, B > 0$. From (\ref{eosexp}) we can immediately read off the
phase diagram in the $\kappa -\beta$ plane. This is shown in
Fig.~\ref{criticalfine} for $N_t=12$. Similar diagrams hold for
different values of $N_t$. For $t > 0$ the chiral condensate vanishes
in the chiral limit. For $t < 0$, on the other hand, chiral symmetry is
spontaneously broken for all $\kappa$ values and 
\begin{equation}
\hat{\sigma}^{\frac{1}{\beta}} = - \frac{B}{A} \, t \, > \, 0
\end{equation}
in the chiral limit. The critical temperature $T_c(m=0)$ is obtained
from $N_t$ and the corresponding lattice spacing at $t=0$.

\section{Chiral susceptibility and Maxwell relation}

The chiral condensate is related to the average plaquette $P$ by means of a
Maxwell relation~\cite{Gockeler}. The plaquette can be computed with high
precision  
and so lends itself to an accurate determination of the critical temperature
of the chiral transition.

Both the chiral condensate and the plaquette can be found from the partial
derivatives of the partition function $Z$:
\begin{eqnarray}
\frac{1}{V} \left.\frac{\partial}{\partial \, \beta} \ln Z
\right|_{\hat{m}} &=& -6\, P  + 2\,\frac{\partial \, \hat{m}_c}{\partial \,
  \beta}\, \hat{\sigma} - 2\, \frac{\partial \, c_{SW}}{\partial \,
  \beta}\, \hat{\delta}\,,\\ 
\frac{1}{V} \left.\frac{\partial}{\partial \, \hat{m}} \ln Z
\right|_\beta &=& 2\, \hat{\sigma} \,,
\end{eqnarray}
where, temporarily, we have set $\hat{\sigma}=\langle\sigma\rangle$, and 
\begin{equation}
\hat{\delta}=\langle \delta\rangle \,, \quad \delta=\frac{i}{2} \,\frac{a^3}{V}
\sum_x \bar{\psi}(x)\, \sigma_{\mu\nu}\, P_{\mu\nu}(x)\, \psi(x) \,.
\end{equation} 
The second derivative $\partial^2 \ln  Z/\partial \beta\, \partial \hat{m}$ can
be expressed in two different orders, 
\begin{equation} 
\frac{1}{V} \frac{\partial^{\,2}}{\partial \,\beta \,\partial\,
  \hat{m}} \ln Z = 2 
\left.\frac{\partial \,\hat{\sigma}}{\partial \, \beta}\right|_{\hat{m}} = -6
\left.\frac{\partial \,P}{\partial \, \hat{m}}\right|_\beta +
2\,\frac{\partial \, \hat{m}_c}{\partial \, \beta}\,
\left.\frac{\partial \, \hat{\sigma}}{\partial\,
  \hat{m}}\right|_{\beta} - 2\,
\frac{\partial \, c_{SW}}{\partial \, \beta}\, \left.\frac{\partial
  \,\hat{\delta}}{\partial \, \hat{m}}\right|_\beta\,.
\label{max}
\end{equation}
Both $\hat{\sigma}$ and $\hat{\delta}$ are chiral order parameters. In
the following we shall neglect the contribution from the clover term
$\hat{\delta}$, mainly for the sake of clarity. We may do so,
because it is suppressed by two orders of the lattice spacing with respect
to the chiral condensate~\cite{Kremer,Doi,DiGiacomo}:
\begin{equation}
\frac{\hat{\delta}}{\hat{\sigma}} = O(a^2)\,.
\end{equation}
This leaves us with the relation
\begin{equation}
\left.\frac{\partial \,P}{\partial \, \hat{m}}\right|_\beta -
\frac{1}{3} \, \frac{\partial \, \hat{m}_c}{\partial \, \beta}\,
\left.\frac{\partial \, \hat{\sigma}}{\partial\,
  \hat{m}}\right|_{\beta}= - \frac{1}{3}
\left.\frac{\partial \,\hat{\sigma}}{\partial \,
  \beta}\right|_{\hat{m}}  = \frac{1}{3} 
\left.\frac{\partial \,\hat{m}}{\partial \, \beta}\right|_{\hat{\sigma}} \,
\left.\frac{\partial \,\hat{\sigma}}{\partial \,\hat{m}}\right|_\beta \,.
\label{plaquette}
\end{equation}
Above identity is called the Maxwell relation. It holds for any lattice
size and for all values of $\beta$ and $m$. The right-hand form of the
identity is the most useful, as it gives us the chiral susceptibility: 
\begin{equation}
\chi_\sigma \equiv \left.\frac{\partial \, \hat{\sigma}}{\partial\, \hat{m}}
\right|_{\beta} = 3 \left(\left.\frac{\partial \,\hat{m}_c}{\partial \,
      \beta} + \frac{\partial \,\hat{m}}{\partial \,
      \beta}\right|_{\hat{\sigma}}\right)^{-1} \, \left.\frac{\partial
  \,P}{\partial \, 
      \hat{m}}\right|_\beta  = 3 \left(\left.\frac{\partial}{\partial \,
      \beta} \, \frac{1}{2\,\kappa}\,\right|_{\hat{\sigma}}\right)^{-1} \,
\left.\frac{\partial \,P}{\partial \, 
      \hat{m}}\right|_\beta \,.
\label{chiplaq}
\end{equation}
In both correlators, $\chi_\sigma$ and $\left.(\partial
\,P/\partial\,\hat{m})\right|_\beta$, the additive renormalization
$\sigma_0$ drops out, so that we may identify $\hat{\sigma}$ with
$\langle\sigma_{-}\rangle$ again. 

Before we can identify any peak of $\partial P/\partial \hat{m}$ with a peak
of $\chi_\sigma$, we need to make sure that the factor in front of
$\partial P/\partial \hat{m}$, 
\begin{equation}
\mu^{-1} = 3 \left(\left.\frac{\partial \,\hat{m}_c}{\partial \,
      \beta} + \frac{\partial \,\hat{m}}{\partial \,
      \beta}\right|_{\hat{\sigma}}\right)^{-1} \,,
\end{equation}
is `well behaved'. The QCDSF collaboration finds for the first term
\begin{equation}
\frac{\partial \,\hat{m}_c}{\partial \, \beta} \, = \,
\left\{ \begin{array}{c@{\quad}c} -0.1536 & \beta=5.20 \\ -0.1177 & \beta=5.25
  \\ -0.0919 & \beta=5.29 \end{array}\right.                   \,.
\label{first}
\end{equation}
The second term can be derived from the equation of state
(\ref{eosexp}), which gives   
\begin{equation}
\left.\frac{\partial \,\hat{m}}{\partial \, \beta}\right|_{\hat{\sigma}} =
B\, \hat{\sigma}^{\delta-\frac{1}{\beta}} \,.
\label{second}
\end{equation}

\clearpage
\begin{figure}[t]
\vspace*{-1.75cm}
\begin{center}
\epsfig{file=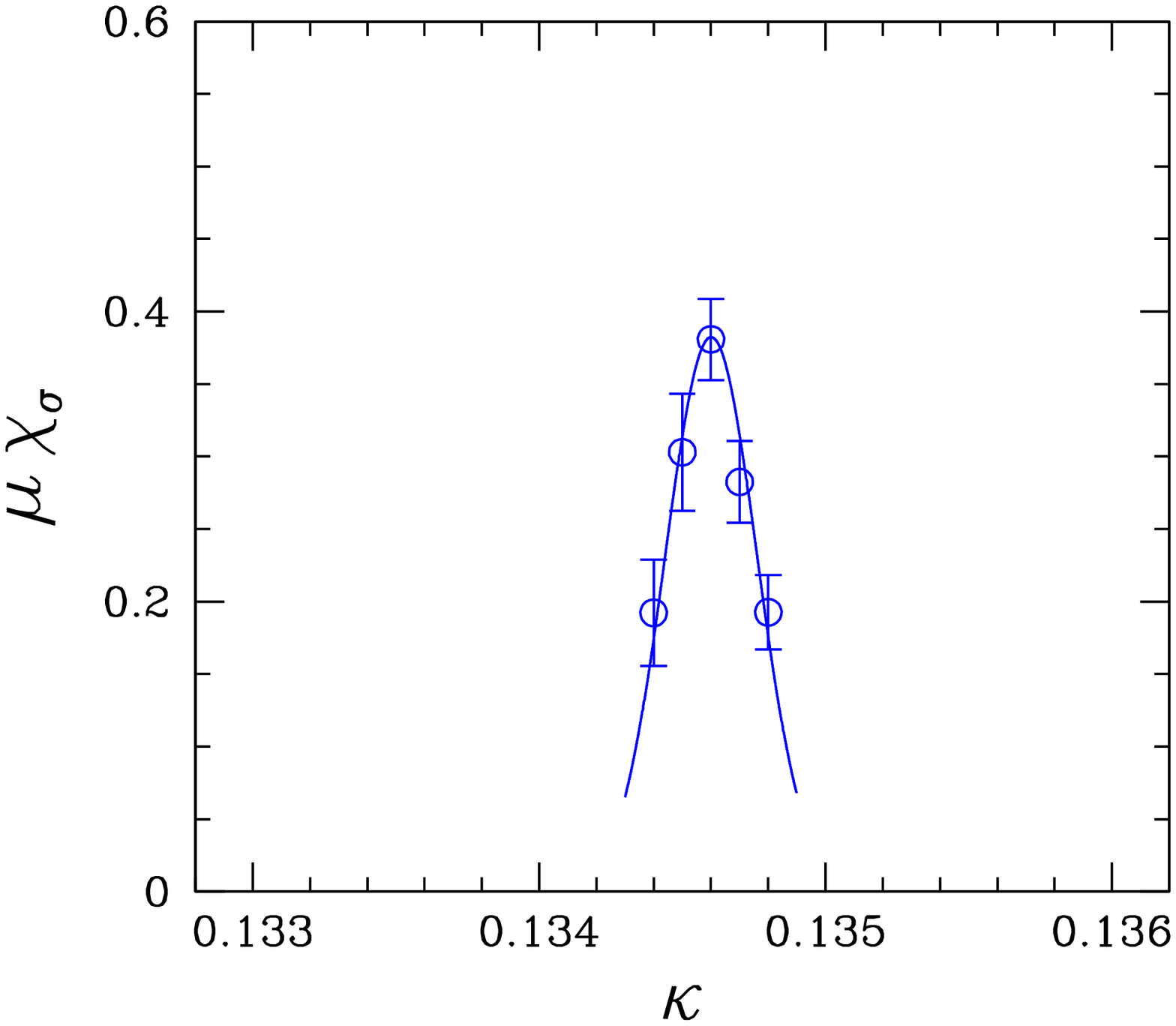,height=8cm,clip=}\\[-4em]
\epsfig{file=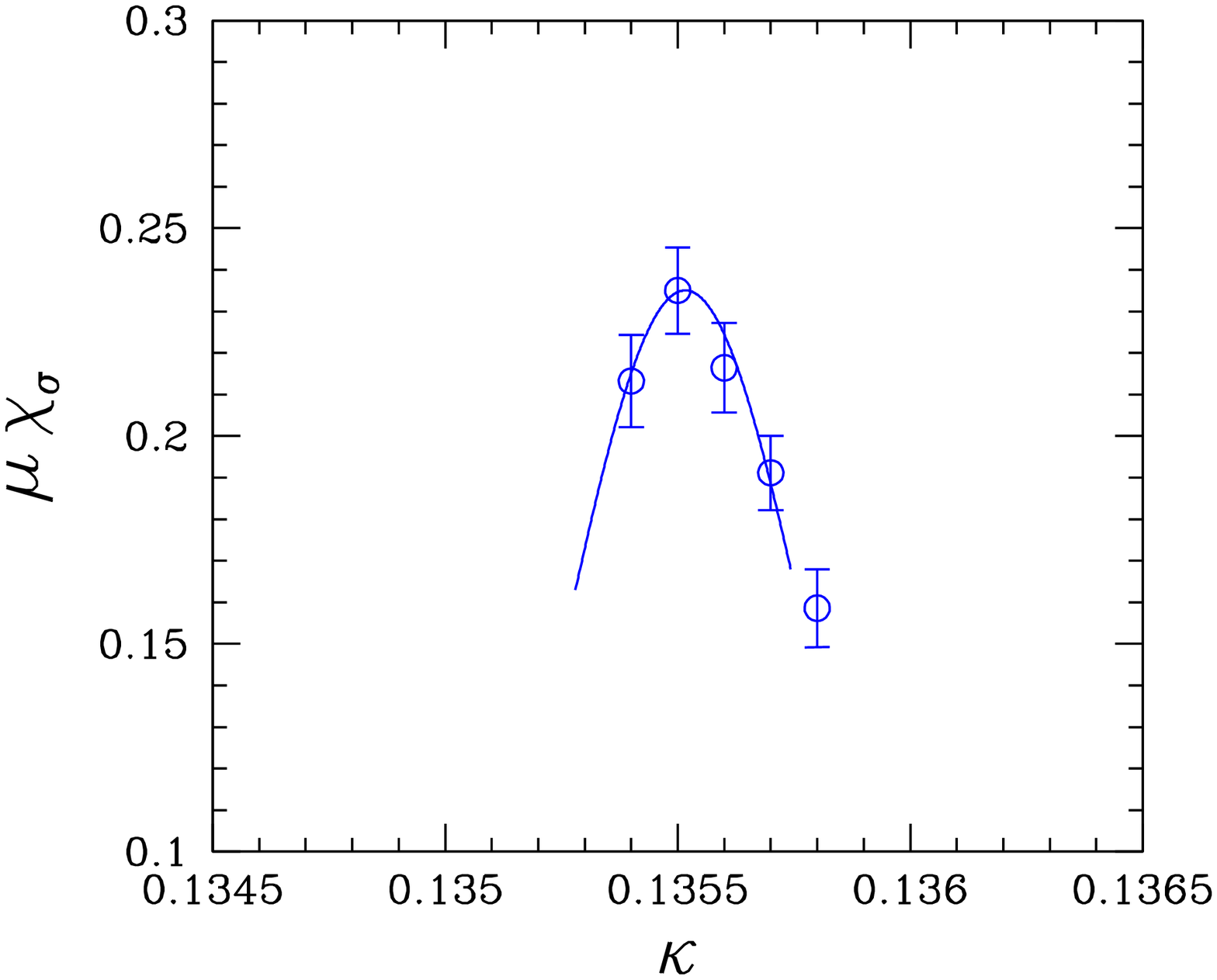,height=8cm,clip=}\\[-4em]
\epsfig{file=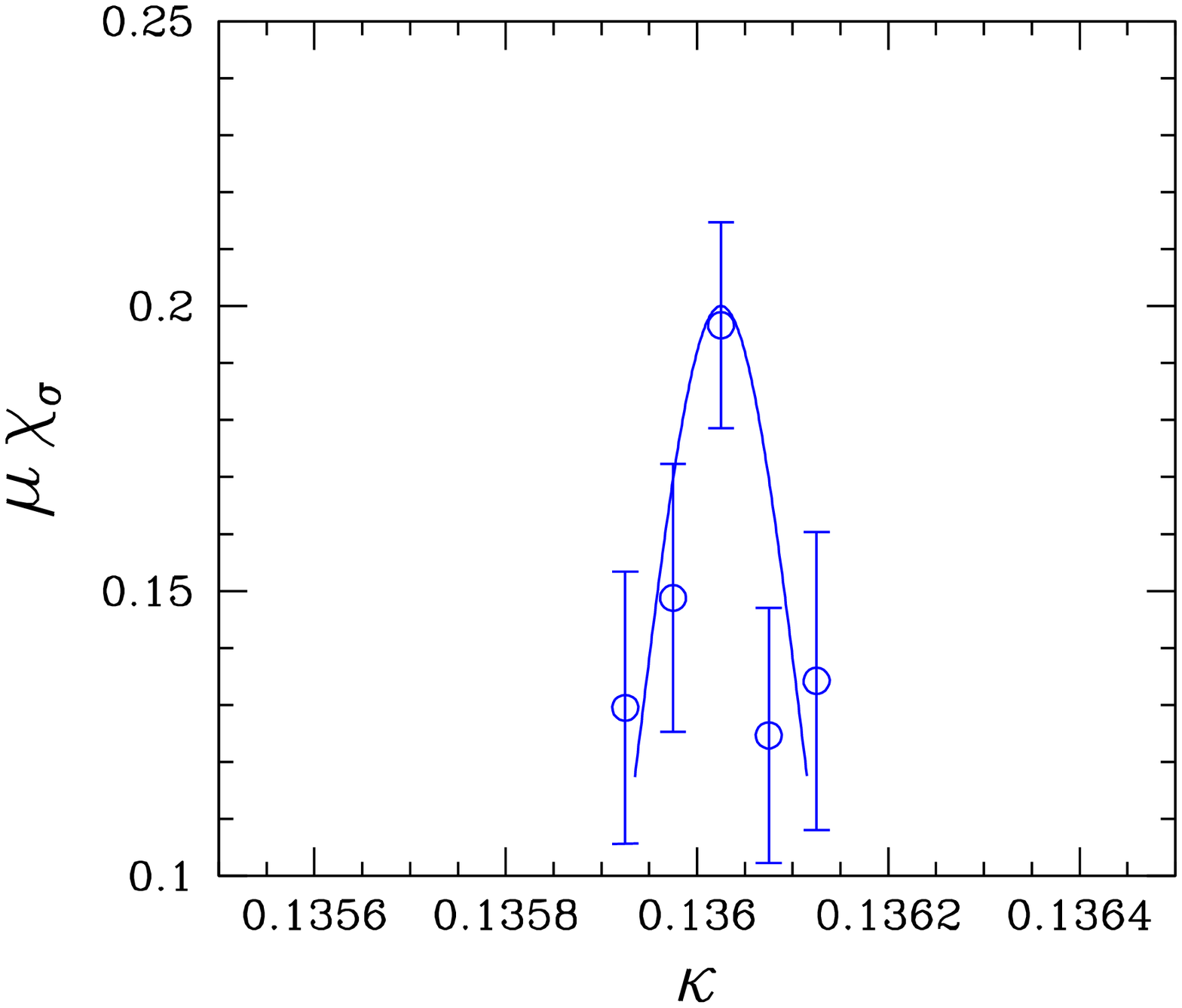,height=8cm,clip=}
\end{center}
\vspace*{-0.5cm}
\caption{The chiral susceptibility on the $16^3\, 8$ (top), $24^3\,
  10$ (middle) and $32^3\, 12$ lattice (bottom) at $\beta=5.20$,
  $5.20$ and $5.25$, respectively, together with a Gaussian.}   
\label{chimax}
\end{figure}
\clearpage
\begin{figure}[t]
\vspace*{-1.75cm}
\begin{center}
\epsfig{file=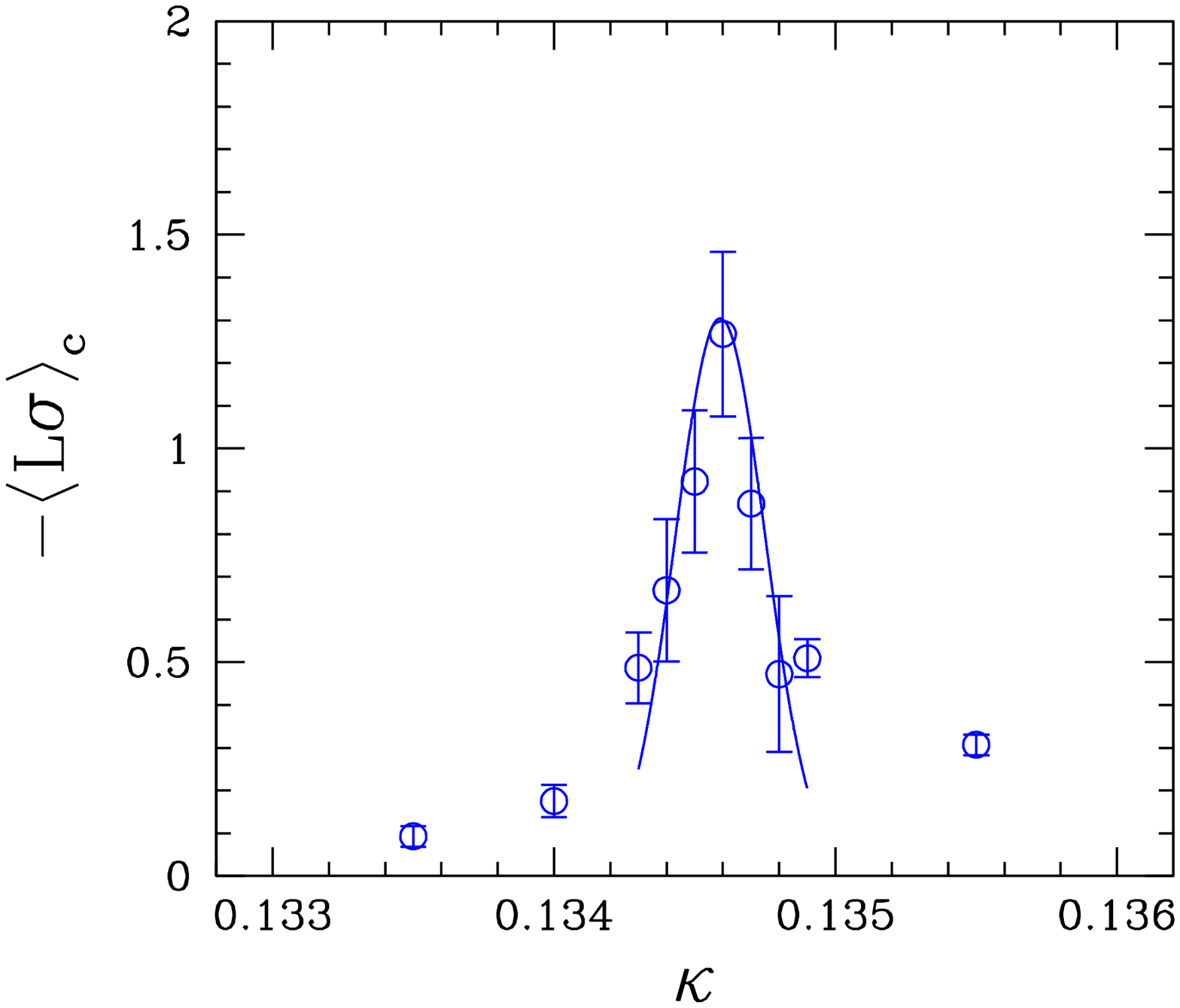,height=8cm,clip=}\\[-4em]
\epsfig{file=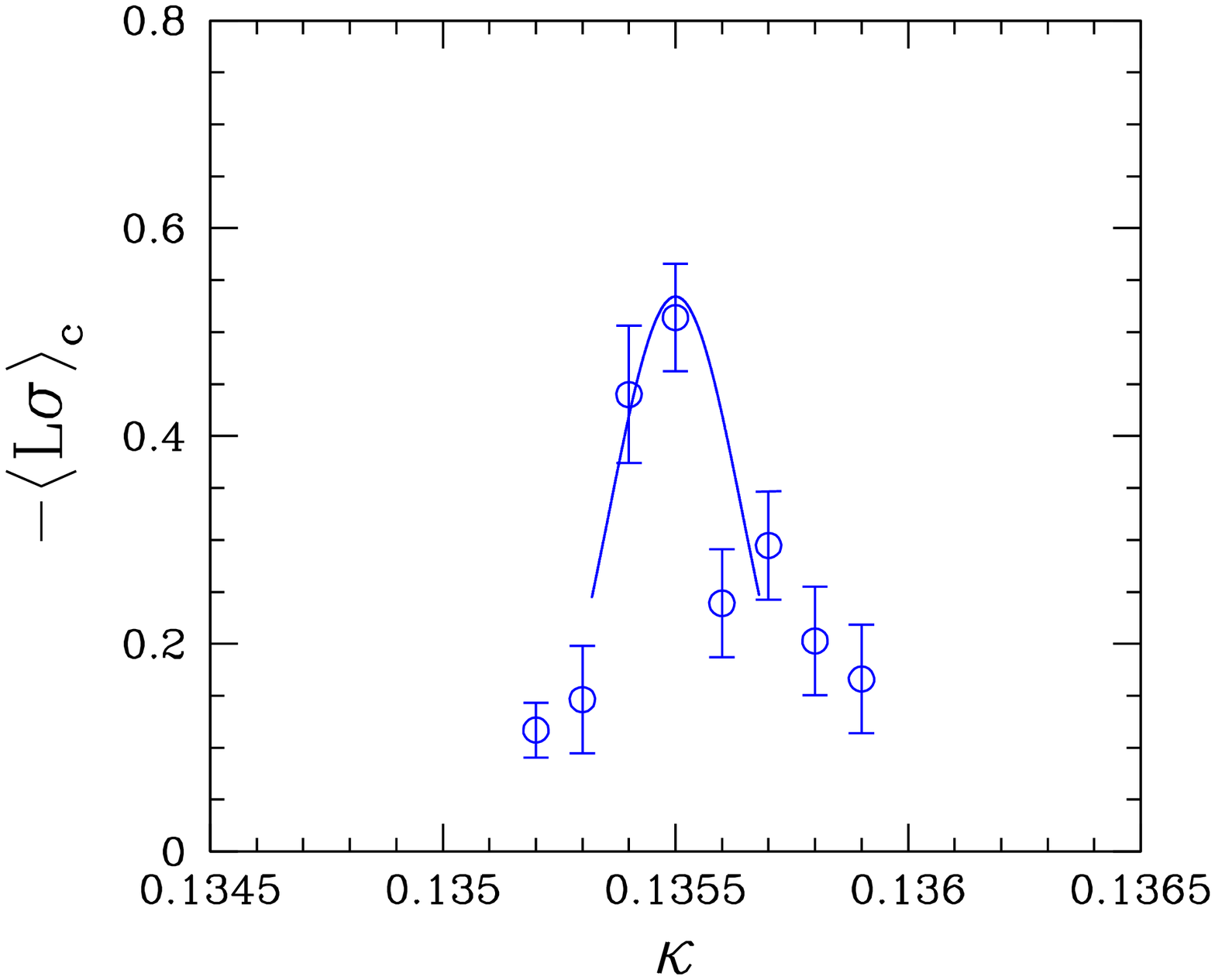,height=8cm,clip=}\\[-4em]
\epsfig{file=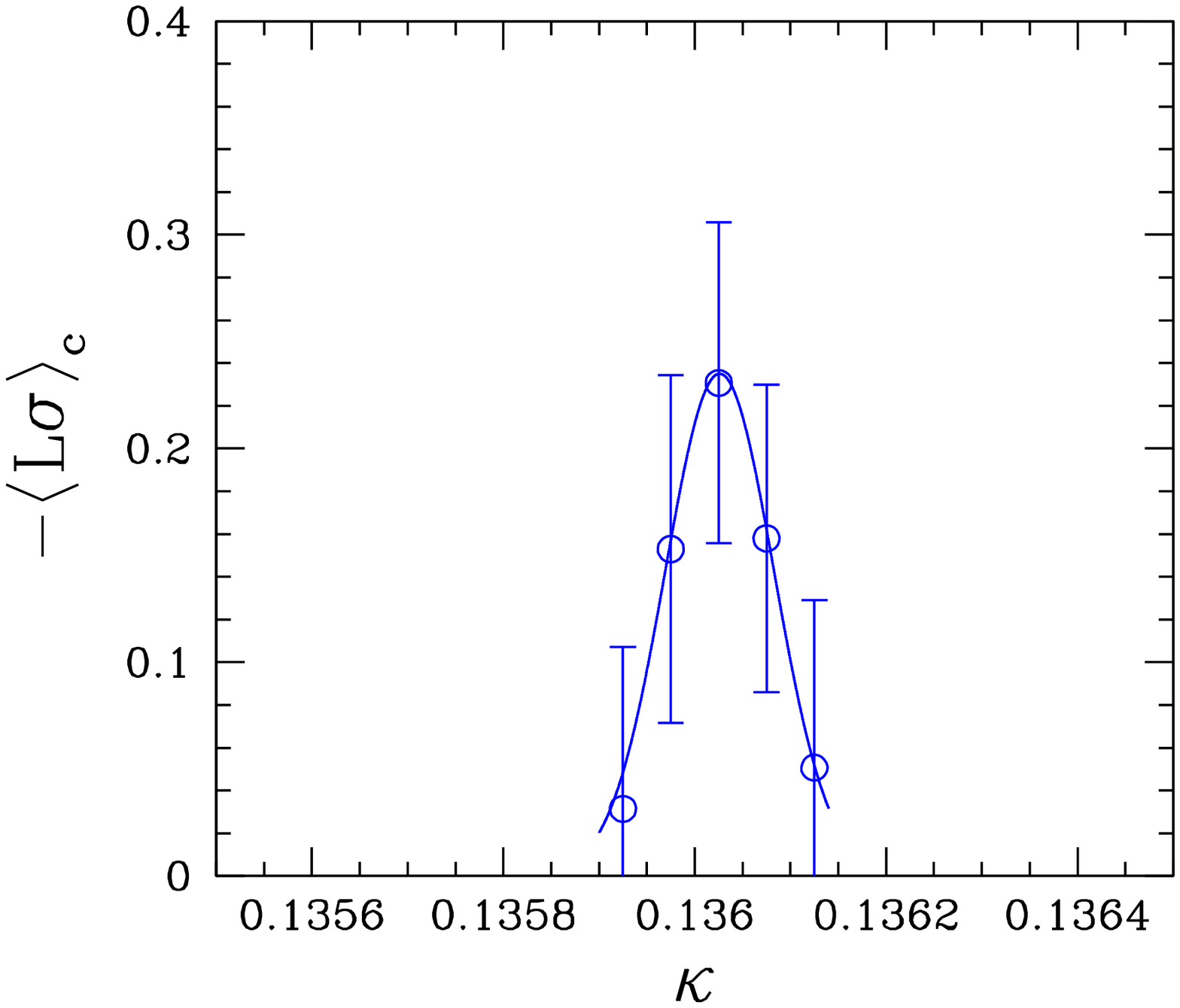,height=8cm,clip=}
\end{center}
\vspace*{-0.5cm}
\caption{The correlator $\langle L\sigma\rangle_c$ on the $16^3\, 8$
  (top), $24^3\, 10$ (middle) and $32^3\, 12$ lattice (bottom) at $\beta=5.20$,
  $5.20$ and $5.25$, respectively, together with a Gaussian.}   
\label{chiL}
\end{figure}

\clearpage
\noindent
It vanishes proportional to $\hat{m}$ in the chiral limit, so that for
small masses $\mu^{-1}$ is dominated by the first term (\ref{first}). 

It does not matter whether we use the renormalized or unrenormalized
subtracted scalar density to determine the transition temperature, because $Z_S$
is an extremely smooth function of $\hat{m}$~\cite{Gockeler2}, which varies by
less than $1\%$ over the transition region.  

The partial differential equation~(\ref{plaquette}) can be solved by 
\begin{equation}
P - P(m=0)= \frac{1}{3}\, \int_0^{\hat{\sigma}} d\sigma^\prime\,
\frac{\partial\, 
  \left[\hat{m}_c + \hat{m}(\sigma^\prime,t)_c\right]}{\partial\, \beta}  \,, 
\label{solution}
\end{equation}
where we have made use of the fact that $\hat{\sigma}=0$ in the chiral limit.
For the {\it ansatz} (\ref{second}) we find
\begin{equation}
P - P(m=0) = \frac{B}{3\, (\delta - \frac{1}{\beta}+1)} \, 
\hat{\sigma}^{\delta-\frac{1}{\beta}+1} + \frac{1}{3}\, \frac{\partial
  \,\hat{m}_c}{\partial \, \beta}\, \hat{\sigma}\,.
\label{plaqeos}
\end{equation}
Knowing the critical exponents and $B$, we may compute the chiral
condensate from the average plaquette, and {\it vice versa}.

The chiral transition has merely a small effect on the average plaquette. The
reason is that only a fraction of a percent of the magnitude of $P$ is of
nonperturbative origin~\cite{Horsley}. Nevertheless, with accurate data
of the plaquette the Maxwell relation proves to be a viable tool to locate the 
position of the chiral phase transition.
We use the symmetric difference quotient method, 
\begin{equation}
f^\prime(x)=\frac{f(x+\epsilon)-f(x-\epsilon)}{2 \epsilon} \,,
\label{numdiff}
\end{equation}
to compute the derivative of the plaquette, $\displaystyle \partial
P/\partial \hat{m}|_\beta$. 
In Fig.~\ref{chimax} we plot the chiral condensate $\mu \chi_\sigma$
as a function of $\kappa$ for three representative lattices, which we chose
to be the same as those in Fig.~\ref{susL}.

\section{Transition temperature(s)}

\begin{figure}[t]
\vspace*{-0.35cm}
\begin{center}
\epsfig{file=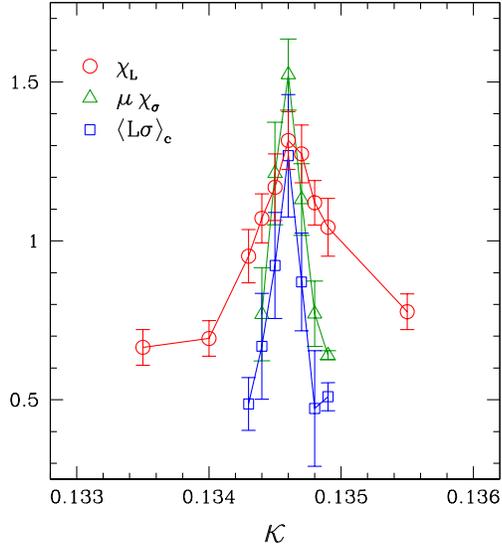,height=8cm,clip=}
\end{center}
\vspace*{-0.75cm}
\caption{The Polyakov-loop susceptibility $\chi_L$ ($\times 7$), the chiral
  susceptibility $\chi_\sigma$ ($\times 4$) and the correlator $\langle
  L\sigma\rangle_c$ on the $16^3\, 8$ lattice at $\beta=5.20$.}   
\label{com}
\end{figure}

To complete the calculation of the susceptibility $\langle
\omega^2\rangle_c$ introduced in (\ref{phase2}), what remains to be
computed is the correlator $\langle L\sigma\rangle_c$. This can be obtained
from the derivative of the average Polyakov loop with respect to mass:
\begin{equation}
\langle L\sigma\rangle_c = \left.\frac{\partial \langle L\rangle}{\partial
    \hat{m}}\right|_\beta \,.
\label{Ls}
\end{equation}
Again, the additive renormalization constant drops out in (\ref{Ls}),
and it does not make any difference whether we use the renormalized
Polyakov loop~\cite{Kaczmarek} or not, because the renormalization factor is
basically a function of $\beta$ only, being an ultraviolet quantity.
In Fig.~\ref{chiL} we plot $\langle L\sigma\rangle_c$ on the lattices
of Figs.~\ref{susL} and \ref{chimax}, where we have expressed the
derivative by the symmetric 
difference (\ref{numdiff}) as before. Not only do we find that the 
Polyakov loop and the chiral condensate are strongly correlated within a
narrow temperature range, but
also that the position of the peaks of all three quantities, $\chi_L$,
$\chi_\sigma$ and $\langle L\sigma\rangle_c$, coincide within small
error bars. This is illustrated once more in Fig.~\ref{com} for the $16^3\, 8$
lattice at $\beta=5.20$. 

\begin{figure}[b!]
\vspace*{-2cm}
\begin{center}
\epsfig{file=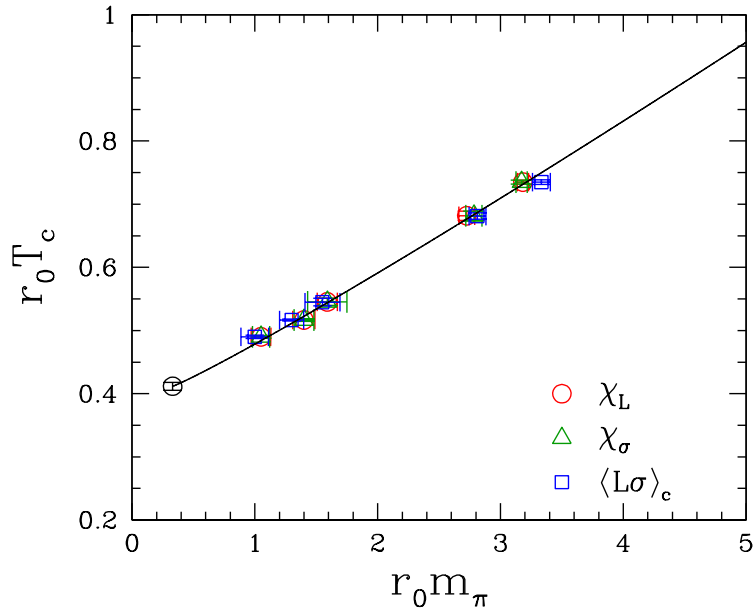,height=10.5cm,clip=}
\end{center}
\vspace*{-0.5cm}
\caption{The pseudocritical temperature $T_c(m)$ as a function of pion
  mass, together with a fit to the power $m_\pi^{1.07}$,
  according to the three-dimensional O(4) model.}   
\label{tcm}
\end{figure}

We may express the $\kappa$ values by the corresponding pion masses at zero
temperature. The latter are known from simulations of the QCDSF
collaboration. In Table~\ref{kt} we give the pseudocritical
temperature $T_c(m)$ and the corresponding pseudocritical pion masses,
$m_\pi^{T_c}$, obtained from the peak of the Polyakov-loop
\begin{table}[t]
\begin{center}
\vspace*{0.25cm}
\begin{tabular}{|c|c|c|c|c|c|}\hline
 & & & \multicolumn{3}{c|}{$r_0\, m_\pi^{T_c}$} \\ 
\raisebox{10pt}{$\beta$} & \raisebox{10pt}{$V$} & \raisebox{10pt}{$r_0\,T_c(m)$} & $\chi_L$ & $\chi_\sigma$ & $\langle
  L\sigma\rangle_c$ \\
\hline
5.20  & $16^3\, 8$\phantom{0} & 0.682(7) & 2.73(6)\phantom{0} & 2.78(6)\phantom{0} & 2.81(7)\phantom{0} \\
5.20  & $24^3\, 10$ & 0.545(6) & 1.59(8)\phantom{0} & 1.59(16) & 1.55(14) \\
5.25  & $24^3\, 8$\phantom{0} & 0.735(3) & 3.18(4)\phantom{0} & 3.17(4)\phantom{0} & 3.33(7)\phantom{0} \\
5.25  & $32^3\, 12$ & 0.490(2) & 1.00(11)& 1.05(8)\phantom{0} & 1.05(7)\phantom{0} \\
5.29  & $24^3\, 12$ & 0.517(2) & 1.49(8)\phantom{0} & 1.40(9)\phantom{0} & 1.3(1)\phantom{00}\\\hline 
\end{tabular}
\end{center}
\caption{The pseudocritical temperatures and corresponding pion
  masses obtained from the peak of $\chi_L$, $\chi_\sigma$ and $\langle
  L\sigma\rangle_c$ on our various lattices.} 
\label{kt}
\end{table}
susceptibility, the chiral susceptibility and the correlator
(\ref{Ls}) of $L$ and $\sigma$, respectively, and in Fig.~\ref{tcm} we
plot the results together with an extrapolation to the chiral
limit. On all lattices the individual pion masses $\displaystyle
m_\pi^{T_c}$ are found to coincide with each other within the error
bars. 
We do not observe any scaling violations. Our complementary runs at
zero temperature did not reveal any visible scaling violations
either. The spatial volume varies from $L_s (\equiv a N_s)=1.4\,\mbox{fm}$ to
$L_s=2.5\,\mbox{fm}$, while the aspect ratio $N_s/N_t$ varies between
$2$ and $3$. Remarkably, all our results fall on a single curve. In
addition, we do not see any finite size effect in direct comparison
at $N_t=8$.

The temperature 
$T_c(m)$ shows an almost linear behavior in the pion mass, in
accord with the prediction (\ref{o4}) of the O(4) model. We thus may
fit the data by the {\it ansatz}
\begin{equation}
T_c(m) = C + D\, (r_0\,m_\pi)^{1.07} \,.
\end{equation}
The result is shown by the solid curve. Setting the scale by the
nucleon mass, the QCDSF collaboration finds $r_0=0.467(15)\,
\mbox{fm}$. Using this value, we obtain at the physical pion mass
\begin{equation}
r_0\,T_c = 0.412(6) \; , \quad T_c=174(3)(6) \, \mbox{MeV} \,,
\end{equation}
where the first error on $T_c$ is statistical, and the second error reflects the
uncertainty in setting the scale. This result is in good agreement with the
deconfining transition 
temperature found by the Wuppertal group, but lies significantly below
the result of the Brookhaven/Bielefeld collaboration.

\section{Conclusions and outlook}

We have simulated QCD at finite temperature with two dynamical flavors
of nonperturbatively improved Wilson fermions on lattices as large as
$N_t=12$ and lattice spacings as low as $0.075\, \mbox{fm}$. The
transition temperature has been computed from the Polyakov-loop
susceptibility, the chiral susceptibility as well as the correlator of
Polyakov loop and chiral condensate. All three temperatures are found
to coincide with each other within the error bars. Our results do not
support the claim of the Wuppertal group~\cite{W}, albeit for
$N_f=2+1$ flavors, that the deconfining and  
chiral transitions take place at distinctly separated temperatures. The
critical behavior appears to be in accord with the predictions of the O(4)
Heisenberg model, at least as far as the quark mass dependence of
$T_c$ is concerned, while a first order transition~\cite{Cossu} is
very unlikely. However, further simulations at smaller quark masses
are needed in order to confirm this conclusion beyond doubts. 

The Maxwell relation has proven to be a powerful tool in unveiling the
phase structure of clover fermions. It would be interesting to test
(\ref{solution}), and the equation of state (\ref{eos}) itself, by
direct calculation of the chiral condensate.
As we already mentioned, the clover term $\hat{\delta}$ is forbidden
by chiral symmetry just like $\hat{\sigma}$. So there is actually no need to
neglect it (as we did) if we are just looking for a quantity
which peaks at the chiral transition.

The next step is to extend the simulations to physical quark
masses. Such calculations require lattices with temporal extent $N_t=14$ and
larger, given the fact that simulations at couplings below $\beta =
5.20$ are not feasible. This requires a major increase in computing
resources. Preliminary investigations at $\beta=5.25$ indicate the
existence of a transition below $\kappa_c$, resulting in $r_0
T_c(m)=0.420(2)$. This makes us feel confident that the curve in
Fig.~\ref{tcm} will not level off towards the staggered result~\cite{BB2}.

\section*{Acknowledgment}

We like to thank the computer centers at KEK (under the Large Scale
Simulation Program No. 07-14-B), JSSC (Moscow), SC MSU
(Moscow), RIKEN and HLRN (Berlin and Hannover) for their generous
allocation of computer time and technical support. Furthermore, we
like to thank Thomas Streuer and Hinnerk St\"uben for help with the HMC code.
VGB, SMM and MIP are supported in part through grants RFBR~08-02-00661,
RFBR~09-02-0033, DFG-RFBR~436~RUS and NSH-679.2008.2, and like to
thank their colleagues at RFNC-VNIIEF for invaluable help with the
computations.  

\clearpage

\section*{\hspace*{-0.65cm} \bf APPENDIX: TABLES OF RESULTS}

\begin{table}[h]
\begin{center}
\vspace*{1.5cm}
\begin{tabular}{|c|c|c|c|c|c|c|c|}\hline
\multicolumn{8}{|c|}{$\beta=5.20$} \\ \hline\hline
\multicolumn{4}{|c|}{$16^3\, 8$}&\multicolumn{4}{c|}{$24^3\,10$}
\\ \hline
$\kappa$ & $P$ & $\langle L\rangle$ & $\chi_L$ &
$\kappa$ & $P$ & $\langle L\rangle$ & $\chi_L$ \\ \hline
0.1330 & &0.0016(3) &0.075(3)\phantom{0} &0.1348 &0.467177(33) &0.0017(2) & 0.066(4)\\ \hline
0.1335 & &0.0026(6) &0.095(8)\phantom{0} &0.1352 &0.464857(63) &0.0027(2) & 0.065(4)\\ \hline
0.1340 & &0.0042(6) &0.099(8)\phantom{0} &0.1353 &0.464367(38) &0.0033(3) & 0.074(6)\\ \hline
0.1343 &0.469500(166) &0.0065(6) &0.136(12) &0.1354 &0.463854(38) &0.0035(2) & 0.080(4)\\ \hline
0.1344 &0.468949(189) &0.0096(7) &0.153(11) &0.1355 &0.463204(47) &0.0057(2) & 0.094(5)\\ \hline
0.1345 &0.468436(116) &0.0102(7) &0.167(15) &0.1356 &0.462574(42) &0.0063(2) & 0.082(4)\\ \hline
0.1346 &0.467274(118) &0.0147(6) &0.188(13) &0.1357 &0.462027(35) &0.0070(2) & 0.082(3)\\ \hline
0.1347 &0.466334(102) &0.0172(8) &0.182(13) &0.1358 &0.461536(24) &0.0079(2) & 0.078(3)\\ \hline
0.1348 &0.465717(101) &0.0195(6) &0.160(10) &0.1359 &0.461167(37) &0.0081(2) & 0.075(4)\\ \hline
0.1349 &0.465274(98)\phantom{0} &0.0198(6) &0.149(13) &0.1360 &0.460670(36) &0.0088(2) & 0.077(6)\\ \hline
0.1355 & &0.0293(6) &0.111(8)\phantom{0} & & & & \\ \hline
0.1360 &0.460441(57)\phantom{0} &0.0290(4) &0.121(7)\phantom{0} & & & & \\ \hline
\end{tabular}
\end{center}
\vspace*{1.0cm}
\caption{The average plaquette, the average Polyakov loop and the
  Polyakov-loop susceptibility on the $16^3\, 8$ and $24^3\, 10$
  lattice at $\beta=5.20$ against $\kappa$. The data is based on
  $O(10,000)$ trajectories at $\kappa$ values in the immediate
  vicinity of the peak of the Polyakov-loop susceptibility and on
  $O(5,000)$ trajectories towards the edges. The numbers refer to
  trajectory lengths of unit one.}  
\label{tab520}
\end{table}

\clearpage

\begin{table}[h]
\begin{center}
\vspace*{1.25cm}
\begin{tabular}{|c|c|c|c|c|c|c|c|}\hline
\multicolumn{8}{|c|}{$\beta=5.25$} \\ \hline\hline
\multicolumn{4}{|c|}{$16^3\, 8$}&\multicolumn{4}{c|}{$24^3\,8$}
\\ \hline
$\kappa$ & $P$ & $\langle L\rangle$ & $\chi_L$ &
$\kappa$ & $P$ & $\langle L\rangle$ & $\chi_L$ \\ \hline
0.1330\phantom{0} &0.46879(10) &0.0025(7)\phantom{0} &0.07(1)\phantom{00} &0.1340 &0.46326(7)\phantom{0} &0.0090(4)\phantom{0} & 0.144(14)\phantom{0}\\ \hline
0.1337\phantom{0} & &0.0057(8)\phantom{0} &0.158(20) &0.1341 &0.46258(6)\phantom{0} &0.0127(4)\phantom{0} & 0.160(11)\phantom{0}\\ \hline
0.13375 &0.46454(10) &0.0065(6)\phantom{0} &0.151(10) &0.1342\phantom{0} &0.46197(6)\phantom{0} &0.0148(5)\phantom{0} & 0.205(22)\phantom{0}\\ \hline
0.1339\phantom{0} &0.46370(11) &0.0084(7)\phantom{0} &0.173(19) &0.1343\phantom{0} &0.46151(7)\phantom{0} &0.0164(5)\phantom{0} & 0.231(19)\phantom{0}\\ \hline
0.1340\phantom{0} &0.46315(9\phantom{0}) &0.0105(7)\phantom{0} &0.170(16) &0.1344\phantom{0} &0.46080(6)\phantom{0} &0.0188(5)\phantom{0} & 0.224(19)\phantom{0}\\ \hline
0.1341\phantom{0} &0.46267(10) &0.0115(7)\phantom{0} &0.190(18) &0.1345\phantom{0} &0.46030(4)\phantom{0} &0.0205(3)\phantom{0} & 0.199(17)\phantom{0}\\ \hline
0.1342\phantom{0} &0.46188(9)\phantom{0} &0.0145(6)\phantom{0} &0.203(19) &0.1346\phantom{0} &0.45985(10) &0.0220(4)\phantom{0} & 0.172(16)\phantom{0}\\ \hline
0.13425 &0.46152(8)\phantom{0} &0.0167(7)\phantom{0} &0.229(19) &\multicolumn{4}{c|}{$32^3\, 12$}\\ \hline
0.1343\phantom{0} &0.46136(8)\phantom{0} &0.0164(5)\phantom{0} &0.207(11) &$\kappa$ & $P$ & $\langle L\rangle$ & $\langle L^2\rangle_c$ \\ \hline
0.1344\phantom{0} &0.46062(12) &0.0182(8)\phantom{0} &0.200(14) & 0.1359\phantom{0} &0.45608(2)\phantom{0} &0.00193(6) &0.0626(13) \\ \hline
0.1345\phantom{0} &0.46044(26) &0.0201(7)\phantom{0} &0.221(17)\phantom{0} & 0.13595&0.45591(3)\phantom{0} &0.00197(8) &0.0613(20) \\ \hline
0.1350\phantom{0} &0.46039(16) &0.0267(10) &0.163(17) & 0.1360\phantom{0}&0.45571(2)\phantom{0} &0.00218(7) &0.0627(18) \\ \hline
0.1361\phantom{0} &0.45479(8)\phantom{0} &0.0328(4)\phantom{0} &0.144(8)\phantom{0} & 0.13605&0.45544(2)\phantom{0} &0.00249(7) &0.0649(17) \\ \hline
& & & &0.1361\phantom{0} &0.45527(2)\phantom{0} &0.00270(6) &0.0634(23) \\ \hline
& & & &0.13615 &0.45509(3)\phantom{0} &0.00277(8) &0.0631(20) \\ \hline
\end{tabular}
\end{center}
\vspace*{1.0cm}
\caption{The average plaquette, the average Polyakov loop and the
  Polyakov-loop susceptibility on the $16^3\, 8$, $24^3\, 8$ and
  $32^3\, 12$ lattice at $\beta=5.25$ against $\kappa$. The data is
  based on $O(10,000 - 15,000)$ ($O(10,000)$) trajectories at $\kappa$
  values in the immediate vicinity of the peak of the Polyakov-loop
  susceptibility and on $O(2,000 - 4,000)$ ($O(3,000 - 5,000)$)
  trajectories right at the edges on the $16^3\, 8$ ($24^3\, 8$)
  lattice. On the $32^3\, 12$ lattice we have accumulated $O(3,000 - 7,000)$
  trajectories at the individual $\kappa$ values.} 
\label{tab525}
\end{table}

\clearpage

\begin{table}[h]
\begin{center}
\vspace*{5.25cm}
\begin{tabular}{|c|c|c|c|}\hline
\multicolumn{4}{|c|}{$\beta=5.29\quad 24^3\, 12$} \\ \hline\hline
$\kappa$ & $P$ & $\langle L\rangle$ & $\chi_L$ \\ \hline
0.1357 &0.45223(4) &0.00165(15) &0.0665(38)\phantom{0} \\ \hline
0.1358 &0.45185(2) &0.00184(10) &0.0667(18)\phantom{0} \\ \hline
0.1359 &0.45153(2) &0.00238(11) &0.0695(21)\phantom{0} \\ \hline
0.1360 &0.45114(3) &0.00272(11) &0.0684(19) \\ \hline
0.1361 &0.45086(4) &0.00340(11) &0.0623(27) \\ \hline
\end{tabular}
\end{center}
\vspace*{1.0cm}
\caption{The average plaquette, the average Polyakov loop and the
  Polyakov-loop susceptibility on the $24^3\, 12$ lattice at
  $\beta=5.29$ against $\kappa$. The data is based on $O(5,000)$
  trajectories at the central $\kappa$ values and on $O(2,000 - 3,000)$
  trajectories at the outer $\kappa$ values.} 
\label{tab529}
\end{table}


\clearpage

\begin{thebibliography}{99}
\frenchspacing

\bibitem{BB}
  M.~Cheng, N.~H.~Christ, S.~Datta, J.~van der Heide, C.~Jung, F.~Karsch,
  O.~Kaczmarek, E.~Laermann, R.~D.~Mawhinney, C.~Miao, P.~Petreczky, K.~Petrov,
  C.~Schmidt and T.~Umeda,
  Phys.\ Rev.\  D {\bf 74}, 054507 (2006)
  [arXiv:hep-lat/0608013].

\bibitem{W}
  Y.~Aoki, Z.~Fodor, S.~D.~Katz and K.~K.~Szabo,
  Phys.\ Lett.\  B {\bf 643}, 46 (2006)
  [arXiv:hep-lat/0609068].

\bibitem{BB2}
  M. Cheng, N.~H. Christ, S. Datta, J. van der Heide, C. Jung,
  F. Karsch, O. Kaczmarek, E.~Laermann, R.~D. Mawhinney, C. Miao,
  P. Petreczky, K. Petrov, C. Schmidt, W. Soeldner and T. Umeda, 
  Phys.\ Rev.\  D {\bf 77}, 014511 (2008)
  [arXiv:0710.0354 [hep-lat]].

\bibitem{W2}
  Y.~Aoki, S.~Borsanyi, S.~D\"urr, Z.~Fodor, S.~D.~Katz, S.~Krieg and
  K.~K.~Szabo, 
  JHEP {\bf 0906}, 088 (2009)
  [arXiv:0903.4155 [hep-lat]].

\bibitem{BBB}
  A. Bazavov, T. Bhattacharya, M. Cheng, N.~H. Christ, C. DeTar,
  S. Ejiri, S.~Gottlieb, R.~Gupta, U.~M. Heller, K. Huebner,
  C. Jung, F. Karsch, E. Laermann, L. Levkova, C. Miao,
  R.~D. Mawhinney, P. Petreczky, C. Schmidt, R.~A. Soltz, W. Soeldner,
  R. Sugar, D. Toussaint and P. Vranas, 
  Phys.\ Rev.\  D {\bf 80}, 014504 (2009)
  [arXiv:0903.4379 [hep-lat]].

\bibitem{Weise}
  S.~Roessner, T.~Hell, C.~Ratti and W.~Weise,
  Nucl.\ Phys.\  A {\bf 814}, 118 (2008)
  [arXiv:0712.3152 [hep-ph]].

\bibitem{mix}
  S.~Digal, E.~Laermann and H.~Satz,
  Eur.\ Phys.\ J.\  C {\bf 18}, 583 (2001)
  [arXiv:hep-ph/0007175].

\bibitem{mix2}
  S.~Digal, E.~Laermann and H.~Satz,
  Nucl.\ Phys.\  A {\bf 702}, 159 (2002).

\bibitem{mix3}
  K.~Fukushima,
  Phys.\ Lett.\  B {\bf 553}, 38 (2003)
  [arXiv:hep-ph/0209311].

\bibitem{mix4}
  K.~Fukushima,
  Phys.\ Rev.\  D {\bf 68}, 045004 (2003)
  [arXiv:hep-ph/0303225].

\bibitem{DIK}
  V.~G.~Bornyakov, S.~M.~Morozov, Y.~Nakamura, M.~I.~Polikarpov,
  G.~Schierholz and T.~Suzuki,
  PoS {\bf LAT2007}, 171 (2007)
  [arXiv:0711.1427 [hep-lat]].

\bibitem{Rainer}
  K.~Jansen and R.~Sommer,
  Nucl.\ Phys.\  B {\bf 530}, 185 (1998)
  [Erratum-ibid.\  B {\bf 643}, 517 (2002)]
  [arXiv:hep-lat/9803017].

\bibitem{QCDSF}
  A. Ali Khan, T. Bakeyev, M. G\"ockeler, R. Horsley, D. Pleiter,
  P. Rakow, A. Sch\"afer, G.~Schierholz and H. St\"uben,
  Phys.\ Lett.\  B {\bf 564}, 235 (2003)
  [arXiv:hep-lat/0303026];
  A. Ali Khan, T.~Bakeyev, M. G\"ockeler, R. Horsley, D. Pleiter,
  P.~E.~L. Rakow, A. Sch\"afer, G. Schierholz and H. St\"uben, 
  Nucl.\ Phys.\ Proc.\ Suppl.\  {\bf 129}, 853 (2004)
  [arXiv:hep-lat/0309078].

\bibitem{Hasenbusch}
  M.~Hasenbusch,
  Phys.\ Lett.\  B {\bf 519}, 177 (2001)
  [arXiv:hep-lat/0107019].

\bibitem{Sexton}
  J.~C.~Sexton and D.~H.~Weingarten,
  Nucl.\ Phys.\  B {\bf 380}, 665 (1992).

\bibitem{Roger}
  M.~G\"ockeler, R.~Horsley, Y.~Nakamura, H.~Perlt, D.~Pleiter,
  P.~E.~L.~Rakow, G.~Schierholz, A.~Schiller, T.~Streuer, H.~St\"uben
  and J.~M.~Zanotti,
  PoS {\bf LAT2007}, 041 (2007)
  [arXiv:0712.3525 [hep-lat]];
  N. Cundy, M. G\"ockeler, R. Horsley, T. Kaltenbrunner,
  A. D. Kennedy, Y. Nakamura, H. Perlt, D. Pleiter, P. E. L. Rakow,
  A. Sch\"afer, G. Schierholz, A. Schiller, H. St\"uben and
  J. M. Zanotti, 
  Phys.\ Rev.\  D {\bf 79}, 094507 (2009)
  [arXiv:0901.3302 [hep-lat]].

\bibitem{QCDSF2}
  A. Ali Khan, T. Bakeyev, M. G\"ockeler, T. R. Hemmert, R. Horsley,
  A. C. Irving, B. Jo\'{o}, D. Pleiter, P. E. L. Rakow, G. Schierholz and
  H. St\"uben, 
  Nucl.\ Phys.\  B {\bf 689}, 175 (2004)
  [arXiv:hep-lat/0312030];
  M.~G\"ockeler, R.~Horsley, A.~C.~Irving, D.~Pleiter, P.~E.~L.~Rakow,
  G.~Schierholz and H.~St\"uben,
  Phys.\ Lett.\  B {\bf 639}, 307 (2006)
  [arXiv:hep-ph/0409312];
  M.~G\"ockeler, R.~Horsley, A.~C.~Irving, D.~Pleiter, P.~E.~L.~Rakow,
  G.~Schierholz and H.~St\"uben, 
  Phys.\ Rev.\  D {\bf 73}, 014513 (2006)
  [arXiv:hep-ph/0502212];
  M. G\"ockeler, R. Horsley, A.~C. Irving, D. Pleiter, P.~E.~L. Rakow,
  G. Schierholz, H. St\"uben and J.~M. Zanotti,
  Phys.\ Rev.\  D {\bf 73}, 054508 (2006)
  [arXiv:hep-lat/0601004].

\bibitem{cppacs}
  A. Ali Khan, S. Aoki, R. Burkhalter, S. Ejiri, M. Fukugita,
  S. Hashimoto, N. Ishizuka, Y.~Iwasaki, K. Kanaya, T. Kaneko,
  Y. Kuramashi, T. Manke, K.-I. Nagai, M. Okamoto, M.~Okawa,
  H.~P. Shanahan, Y. Taniguchi, A. Ukawa and T. Yoshi\'{e}, 
  Phys.\ Rev.\  D {\bf 64}, 074510 (2001)
  [arXiv:hep-lat/0103028].

\bibitem{PW}
  R.~D.~Pisarski and F.~Wilczek,
  Phys.\ Rev.\  D {\bf 29}, 338 (1984).

\bibitem{cex}
  K.~Kanaya and S.~Kaya,
  Phys.\ Rev.\  D {\bf 51}, 2404 (1995)
  [arXiv:hep-lat/9409001];
  P.~Butera and M.~Comi,
  Phys.\ Rev.\  B {\bf 52}, 6185 (1995)
  [arXiv:hep-lat/9505027];
  H.~G.~Ballesteros, L.~A.~Fernandez, V.~Martin-Mayor and A.~Munoz Sudupe,
  Phys.\ Lett.\  B {\bf 387}, 125 (1996)
  [arXiv:cond-mat/9606203].

\bibitem{DeTar}
  C.~DeTar and U.~M.~Heller,
  Eur.\ Phys.\ J.\  A {\bf 41}, 405 (2009)
  [arXiv:0905.2949 [hep-lat]].

\bibitem{Mendes}
  T.~Mendes,
  PoS {\bf LAT2007}, 208 (2007)
  [arXiv:0710.0746 [hep-lat]].

\bibitem{Kogut}
  J.~B.~Kogut and D.~K.~Sinclair,
  Phys.\ Rev.\  D {\bf 73}, 074512 (2006)
  [arXiv:hep-lat/0603021].

\bibitem{Ejiri}
  S.~Ejiri {\it et al.},
  Phys.\ Rev.\  D {\bf 80}, 094505 (2009)
  [arXiv:0909.5122 [hep-lat]].

\bibitem{Iwasaki}
  Y.~Iwasaki, K.~Kanaya, S.~Kaya and T.~Yoshi\'{e},
  Phys.\ Rev.\ Lett.\  {\bf 78}, 179 (1997)
  [arXiv:hep-lat/9609022].

\bibitem{AAK}
  A.~Ali~Khan, S.~Aoki, R.~Burkhalter, S.~Ejiri, M.~Fukugita,
  S.~Hashimoto, N.~Ishizuka, Y.~Iwasaki, K.~Kanaya, T.~Kaneko,
  Y.~Kuramashi, T.~Manke, K.~Nagai, M.~Okamoto, M.~Okawa, A.~Ukawa,
  T.~Yoshi\'{e}, 
  Phys.\ Rev.\  D {\bf 63}, 034502 (2001)
  [arXiv:hep-lat/0008011].

\bibitem{GS}
  QCDSF collaboration, work in progress.

\bibitem{Gockeler}
  M.~G\"ockeler, R.~Horsley, V.~Linke, P.~E.~L.~Rakow, G.~Schierholz and
  H.~St\"uben, 
  Nucl.\ Phys.\  B {\bf 487}, 313 (1997)
  [arXiv:hep-lat/9605035].

\bibitem{Kremer}
  M.~Kremer and G.~Schierholz,
  Phys.\ Lett.\  B {\bf 194}, 283 (1987).

\bibitem{Doi}
  T.~Doi, N.~Ishii, M.~Oka and H.~Suganuma,
  Prog.\ Theor.\ Phys.\ Suppl.\  {\bf 151}, 161 (2003)
  [arXiv:hep-lat/0303014].

\bibitem{DiGiacomo}
  A.~Di Giacomo and Yu.~A.~Simonov,
  Phys.\ Lett.\  B {\bf 595}, 368 (2004)
  [arXiv:hep-ph/0404044].

\bibitem{Gockeler2}
  M.~G\"ockeler, R.~Horsley, A.~C.~Irving, D.~Pleiter, P.~E.~L.~Rakow,
  G.~Schierholz and H.~St\"uben,
  Phys.\ Lett.\  B {\bf 639}, 307 (2006)
  [arXiv:hep-ph/0409312].

\bibitem{Horsley}
  R.~Horsley, P.~E.~L.~Rakow and G.~Schierholz,
  Nucl.\ Phys.\ Proc.\ Suppl.\  {\bf 106}, 870 (2002)
  [arXiv:hep-lat/0110210].

\bibitem{Kaczmarek}
  O.~Kaczmarek, F.~Karsch, P.~Petreczky and F.~Zantow,
  Phys.\ Lett.\  B {\bf 543}, 41 (2002)
  [arXiv:hep-lat/0207002].

\bibitem{Cossu}
  G.~Cossu, M.~D'Elia, A.~Di Giacomo and C.~Pica,
  PoS {\bf LAT2007}, 219 (2007)
  [arXiv:0710.0174 [hep-lat]].
 


\end{thebibliography}
\end{document}